\documentclass[a4paper,11pt]{article}
\pdfoutput=1 

\usepackage{jcappub} 

\usepackage[T1]{fontenc} 

\title{Breaking a Dark Degeneracy with Gravitational Waves}

\author[a]{Lucas~Lombriser,}
\author[a]{Andy~Taylor}

\affiliation[a]{Institute for Astronomy, University of Edinburgh, Royal Observatory, Blackford Hill, Edinburgh, EH9~3HJ, U.K.}

\emailAdd{llo@roe.ac.uk}
\emailAdd{ant@roe.ac.uk}

\newcommand{\bq}{\begin{equation}}
\newcommand{\eq}{\end{equation}}
\newcommand{\bqa}{\begin{eqnarray}}
\newcommand{\eqa}{\end{eqnarray}}

\newcommand{\hMpc}{h^{-1}{\rm Mpc}}
\newcommand{\rmd}{{\rm d}}
\newcommand{\DL}{D_{\rm L}}

\newcommand{\Om}{\Omega_{\rm m}}
\newcommand{\rhom}{\rho_{\rm m}}
\newcommand{\Deltam}{\Delta_{\rm m}}
\newcommand{\Vm}{V_{\rm m}}

\newcommand{\Sm}{S_{\rm m}}
\newcommand{\delg}{\delta g^{00}}
\newcommand{\Mbar}{\bar{M}}

\newcommand{\aK}{\alpha_{\rm K}}
\newcommand{\aM}{\alpha_{\rm M}}
\newcommand{\aB}{\alpha_{\rm B}}
\newcommand{\aT}{\alpha_{\rm T}}
\newcommand{\aH}{\alpha_{\rm H}}
\newcommand{\cs}{c_{\rm s}}
\newcommand{\cT}{c_{\rm T}}

\abstract{
We identify a scalar-tensor model embedded in the Horndeski action whose cosmological background and linear scalar fluctuations are degenerate with the concordance cosmology.
The model admits a self-accelerated background expansion at late times that is stable against perturbations with a sound speed attributed to the new field that is equal to the speed of light.
While degenerate in scalar fluctuations, self-acceleration of the model implies a present cosmological tensor mode propagation at $\lesssim95\%$ of the speed of light with a damping of the wave amplitude that is $\gtrsim5\%$ less efficient than in general relativity.
We show that these discrepancies are endemic to self-accelerated Horndeski theories with degenerate large-scale structure and are tested with measurements of gravitational waves emitted by events at cosmological distances.
Hence, gravitational-wave cosmology breaks the dark degeneracy in observations of the large-scale structure between two fundamentally different explanations of cosmic acceleration~--~a~cosmological constant and a scalar-tensor modification of gravity.
The gravitational wave event GW150914 recently detected with the aLIGO instruments and its potential association with a weak short gamma-ray burst observed with the Fermi GBM experiment may have provided this crucial measurement.
}

\begin{document}
\maketitle
\flushbottom

\section{Introduction}
%
Determining the nature underlying the late-time accelerated expansion of our Universe remains a difficult puzzle and prime endeavour to cosmologists nearly two decades after its discovery.
In the concordance model, flat $\Lambda$ Cold Dark Matter ($\Lambda$CDM), a cosmological constant $\Lambda$ contributes the bulk of the present energy density in the cosmos and drives the acceleration in accordance with Einstein's theory of general relativity (GR).
The cosmological constant may be attributed to a vacuum energy but its observed value is theoretically not understood, being $\mathcal{O}(10^{60-120})$ smaller than expected.
Alternatively, the presence of a scalar field permeating the Universe and modifying gravity may be responsible for the effect.
However, stringent limitations from experiments within our Solar System, where GR has been verified~\cite{will:05}, must be satisfied.
A number of screening mechanisms~\cite{vainshtein:72,khoury:03a,babichev:09,hinterbichler:10,brax:10,lombriser:14b} have been identified that can suppress gravitational modifications and recover GR locally while still generating significant deviations at cosmological scales.

The nonlinear screening mechanisms can be divided into two main categories: one where screening relies on a suppression of the scalar field value such as is the case for chameleon~\cite{khoury:03a}, symmetron~\cite{hinterbichler:10}, and dilaton~\cite{brax:10} models; the other operating through the gradients of the scalar field such as is the case for models with a Vainshtein~\cite{vainshtein:72} or k-mouflage mechanism~\cite{babichev:09}.
For a model employing the first type of screening to explain cosmic acceleration, the field needs to transition from a large cosmological value to an insignificant one at Solar-System scales.
The known mechanisms have been found to be not sufficiently efficient to accommodate both a self-acceleration and local viability~\cite{hees:11,wang:12}.
A large class of self-accelerated models employing Vainshtein screening~\cite{dvali:00,deffayet:00,nicolis:08,pirtskhalava:15} experienced a similar fate in that they would suffer from theoretical problems~\cite{koyama:07,barreira:13}, yield gravitational dynamics in high-density regions incompatible with Solar-System tests~\cite{barreira:13,jimenez:15}, or be inconsistent with the observed large-scale structure~\cite{fang:08,lombriser:09,barreira:14a}.
Moreover, given the absence of observational evidence of modifications in the large-scale structure with increasingly precise measurements~\cite{planck:15}, these examples suggest that while a nonlinear screening mechanism may be a necessary condition to ensure local viability for scalar-tensor theories with universal coupling of the scalar field to baryons and dark matter, it may potentially not be a sufficient condition.

In a recent Letter~\cite{lombriser:14b}, we have shown the existence of scalar-tensor theories that in the well-tested quasistatic regime of linear cosmological perturbations reproduce the standard Poisson equation, which relates matter to metric fluctuations, and the standard match between the scalar metric potentials of the $\Lambda$CDM model.
Yet in these models the modifications in the gravitational sector may be large enough to give rise to cosmic acceleration.
The allowed model space naturally includes quintessence~\cite{ratra:87,wetterich:87} and k-essence~\cite{armendariz:00} but also introduces previously unknown nonminimally coupled models that are linearly shielded, or cancelled, and can moreover provide an exact match to the $\Lambda$CDM expansion history.
In this paper, adopting the unified dark energy formalism~\cite{creminelli:08,park:10,gubitosi:12,bloomfield:12,bellini:14,gleyzes:14b}, we identify a scalar-tensor model embedded in the Horndeski action~\cite{horndeski:74,deffayet:11,kobayashi:11} that satisfies the linear shielding, or cancellation, conditions~\cite{lombriser:14b}.
It moreover admits a self-acceleration of the late-time expansion of the cosmological background with the linear theory being free of ghost and gradient instabilities and with the sound speed of the new field corresponding to the speed of light.
Solving the modified linearly perturbed Einstein field equations, we show that the scalar fluctuations of the model are degenerate with concordance cosmology at any scale, even beyond the quasistatic regime, but the requirement of self-acceleration leads to a detectable modification in the propagation of the tensor modes.
We show that more generally a deviation between the propagation speed of gravitational waves and the speed of light is symptomatic for self-accelerated Horndeski models recovering the large-scale structure of $\Lambda$CDM or a dark energy model.

In Sec.~\ref{sec:mg}, we briefly review the concept of a self-accelerated late-time expansion due to a modification of gravity.
We discuss the effects that modifying gravity can have in consequence on the formation of structure and propagation of gravitational waves.
We then specialise to scalar-tensor theories and review the linear shielding mechanism.
In Sec.~\ref{sec:model} we present the dark degeneracy that exists between scalar-tensor theories and $\Lambda$CDM in the cosmological background and large-scale structure, for which we analyse the effects on gravitational waves in Sec.~\ref{sec:gw}.
We also comment on the recent gravitational wave detection GW150914 by the Advanced Laser Interferometer Gravitational-Wave Observatory (aLIGO)~\cite{GW150914} and the implication of a potentially associated weak short gamma-ray burst observed with the Fermi Gamma-ray Burst Monitor (GBM)~\cite{GRB150914}.
Note that the model should not be confused with the degeneracy between dark matter and dark energy in an arbitrary joint dark sector fluid pointed out in Ref.~\cite{kunz:07}, although it may be viewed as originating from the same equivalence between the geometric and matter components underlying the Einstein field equations.
Finally, we discuss our results in Sec.~\ref{sec:conclusions}.

\section{Cosmic acceleration from modified gravity} \label{sec:mg}

The lack of a sound theoretical understanding of the late-time accelerated expansion of our Universe motivates the consideration of extensions to Einstein gravity.
However, not every modification of gravity necessarily provides a genuine alternative explanation to the cosmological constant or dark energy.
Conceptually, we may think of a modification of the gravitational interactions as an additional force altering the free-fall geodesic motion of a particle expected from a metric satisfying the Einstein field equations with a modified matter sector.
In comparison to this Einstein frame, we can define a frame in which the particle still moves along geodesics but where the metric satisfies a different field equation with a conventional matter sector.
This is referred to as the Jordan frame.
While in GR, there is no such difference in frame, when gravity is modified relations derived in each frame are distinct in form.

Let us assume that at least for the cosmological background there is a conformal factor $\Omega$ that maps the Jordan frame metric $g_{\mu\nu}$ to the Einstein frame metric
\bq
 \tilde{g}_{\mu\nu}(t,{\bf x}) =  \Omega(t,{\bf x}) \, g_{\mu\nu}(t,{\bf x}) \label{eq:confmap}.
\eq
We assume a four-dimensional statistically spatially homogeneous and isotropic as well as flat ($k_0=0$) universe and define the Friedmann-Lema\^itre-Robertson-Walker (FLRW) metric in the Jordan frame by the line element
\bq
 \rmd s^2 = -\rmd t^2 + a^2(t) \rmd {\bf x}^2 \label{eq:FLRW}
\eq
with scale factor $a(t)$, defining the Hubble parameter $H(t)\equiv\rmd\ln a/\rmd t$, and speed of light in vacuum set to unity here and throughout the paper.
We also apply a transformation of the time coordinate to cast $\tilde{g}_{\mu\nu}$ in the same form as Eq.~(\ref{eq:FLRW}), where $\tilde{a}(\tilde{t})$ is the scale factor in terms of proper time $\tilde{t}$ in the Einstein frame.
We define that a cosmic acceleration in the Jordan frame which is genuinely due to a modification of gravity rather than a dark energy contribution in the matter sector should show no acceleration in $\tilde{a}(\tilde{t})$ (see, e.g., Ref.~\cite{wang:12}).
Hence, while the observation of a late-time accelerated expansion implies
\bq
 \frac{\rmd^2 a}{\rmd t^2} = a\,H^2\left(1+\frac{H'}{H}\right) > 0 \label{eq:accjordan}
\eq
for $a\gtrsim0.6$, for a generic self-acceleration from modified gravity, we expect that
\bqa
 \frac{\rmd^2\tilde{a}}{\rmd \tilde{t}^2} & = & \frac{a\,H^2}{\sqrt{\Omega}} \left[\left(1+\frac{H'}{H}\right)\left(1+\frac{1}{2}\frac{\Omega'}{\Omega}\right) + \frac{1}{2}\left(\frac{\Omega'}{\Omega}\right)' \right] \leq 0, \label{eq:acceinstein}
\eqa
where primes denote derivatives with respect to $\ln a$ here and throughout the paper.
Thus, we expect that $\mathcal{O}(\Omega'/\Omega)\gtrsim1$ for the modification to cause cosmic acceleration at late times.
Note that this requirement excludes modified gravity models where $|\Omega'/\Omega|\ll1$ but where contributions that can be absorbed in the matter sector may serve as an alternative explanation for the late-time acceleration.
In this paper, we focus on models where the effect can predominantly be attributed to a modification of the metric sector and, hence, introduce a considerable difference between the two frames.
Note however that an Einstein frame for a modification in Jordan frame does not always exist for arbitrary metrics~\cite{bettoni:13}.
Here, we only require that the Friedmann equations are mapped into the standard form by a transformation of Eq.~(\ref{eq:FLRW}) with Eq.~(\ref{eq:confmap}).
We shall refer to this relaxed condition as the \emph{Einstein--Friedmann~frame}.

Naturally, the gravitational modification should not only alter the cosmological background evolution but also change structure formation.
To describe the effects on the large-scale structure, we consider linear scalar perturbations around the FLRW metric in the Newtonian gauge with $\Psi\equiv\delta g_{00}/(2g_{00})$ and $\Phi\equiv\delta g_{ii}/(2g_{ii})$.
We adopt the total matter gauge for the matter fluctuations and work in Fourier space.
To simplify notation, the perturbative quantities quoted shall only refer to their Fourier amplitudes of the plane waves with comoving wavenumber $k$, where in linear theory the phases factor out in the field equations.
We restrict to a universe containing only pressureless dust $p_{\rm m}=0$ with background matter density $\rhom$, perturbation $\Deltam$, and large-scale velocity flow $\Vm$.
We then characterise the modifications introduced in the Einstein field equations by an alternative theory of gravity with an effective deviation in the Poisson equation, $\mu$, and the introduction of a gravitational slip, $\gamma$,~\cite{uzan:06,caldwell:07,zhang:07,amendola:07,hu:07b,bertschinger:08,daniel:10}
\bqa
 k_H^2\Psi & = & -\frac{\kappa^2\rhom}{2H^2}\mu(a,k)\Deltam, \label{eq:mu} \\
 \Phi & = & -\gamma(a,k)\Psi, \label{eq:gamma}
\eqa
respectively, where $\kappa^2\equiv8\pi\,G$ with bare gravitational constant $G$ and $k_H\equiv k/(aH)$.
Energy-momentum conservation then closes the system of differential equations for the evolution of the scalar modes with
\bqa
 \Deltam' & = & -k_H\Vm - 3\zeta', \label{eq:encon} \\
 \Vm' & = & -\Vm + k_H\Psi, \label{eq:momcon}
\eqa
where $\zeta=\Phi-\Vm/kH$ is the comoving curvature.
Modifications of gravity also impact the cosmological propagation of gravitational waves described by the linear traceless spatial tensor perturbation, $h_{ij}\equiv g_{ij}/g_{ii}$, of Eq.~(\ref{eq:FLRW}),
\bq
 h_{ij}'' + \left(\nu + 2 + \frac{H'}{H} \right)h_{ij}' + c_{\rm T}^2k_H^2 h_{ij} = 0, \label{eq:gw}
\eq
where $\nu$ parametrises a running of the gravitational coupling that modifies the damping term and $c_{\rm T}$ describes the speed of the tensor mode, which may deviate from the speed of light~\cite{saltas:14}.
Note that a running in the gravitational coupling, in principle, also modifies the evolution of the vector modes~\cite{gleyzes:14b} but this shall not be of concern here as they decay in the models we will consider in Sec.~\ref{sec:model}.

The effective modifications introduced in the evolution of the scalar modes, Eqs.~(\ref{eq:mu}) through (\ref{eq:momcon}), and in the propagation of the tensor modes, Eq.~(\ref{eq:gw}), are sufficiently general to embed Horndeski scalar-tensor gravity~\cite{horndeski:74,deffayet:11,kobayashi:11} and its generalisation to higher-order-derivative equations of motion~\cite{gleyzes:14a}.
Concordance cosmology with standard gravity is restored when $\mu=\gamma=\nu=c_{\rm T}=1$.
In the following, we will specialise our discussion to modifications of gravity linearly equivalent to Horndeski scalar-tensor theory
and describe the corresponding effective parametrisation defined with Eqs.~(\ref{eq:mu}) through (\ref{eq:gw}).

\subsection{Effective field theory} \label{sec:eft}

As an alternative to a cosmological constant, we shall consider the presence of a single low-energy effective scalar field permeating our Universe and causing its late-time expansion to accelerate.
Horndeski gravity~\cite{horndeski:74,deffayet:11,kobayashi:11} describes the most general local, Lorentz-covariant, and four-dimensional scalar-tensor theory where the Euler-Lagrange equations are at most second-order in the derivatives of the scalar and tensor fields.
We adopt the effective field theory (EFT) of cosmic acceleration~\cite{creminelli:08,park:10,gubitosi:12,bloomfield:12,bellini:14,gleyzes:14b}, or unified dark energy formalism, to describe its cosmological background evolution and the linear perturbations around it.
The EFT action for Horndeski gravity, up to quadratic order, reads~\cite{gubitosi:12,bloomfield:12,lombriser:14b}
\bqa
 S & = & \frac{1}{2\kappa^2} \int d^4x\sqrt{-g} \left\{\vphantom{\frac{1^1}{1^1}}\Omega(t) R - 2\Lambda(t) - \Gamma(t) \delg + M_2^4(t) (\delg)^2 - \Mbar_1^3(t) \delg \delta K^{\mu}_{\ \mu} \right. \nonumber\\
 & & \left. - \Mbar_2^2(t) \left[(\delta K^{\mu}_{\ \mu})^2 - \delta K^{\mu}_{\ \nu} \delta K^{\nu}_{\ \mu} - \frac{1}{2} \delg \delta R^{(3)} \right] \right\}
  + \Sm\left[\psi_{\rm m};g_{\mu\nu}\right]. \label{eq:eftaction}
\eqa
The action is written in unitary gauge, where the time coordinate is chosen in order to absorb the scalar field perturbation in the metric $g_{\mu\nu}$.
The scalar-tensor model is then described by a combination of geometric operators that are invariant under time-dependent spatial diffeomorphisms with free time-dependent coefficients, for which we adopted the notation of Ref.~\cite{lombriser:14b}.
$R$ and $R^{(3)}$ denote the four-dimensional and spatial Ricci scalar, respectively, $K_{\mu\nu}$ is the extrinsic curvature tensor, and $n^{\mu}$ describes the normal to surfaces of constant time, where $\delta$ indicates perturbations with respect to the background.
The concordance model is recovered when $\Omega=1$, $\Lambda$ is a constant, and the remaining coefficients vanish, but its phenomenology can be degenerate with other choices of the EFT functions as we will discuss in Sec.~\ref{sec:model} (also see Ref.~\cite{lombriser:14b}).
The Friedmann equations, describing the background evolution, follow from variation of Eq.~(\ref{eq:eftaction}) with respect to the metric~\cite{gubitosi:12,bloomfield:12,lombriser:14b},
\bq
  H^2\left(1 + \frac{\Omega'}{\Omega}\right) = \frac{\kappa^2\rhom + \Lambda + \Gamma}{3\Omega}, \ \ \ \ \ \left(H^2\right)' \left(1 + \frac{1}{2}\frac{\Omega'}{\Omega}\right) + H^2\left(3 + \frac{\Omega''}{\Omega} + 2\frac{\Omega'}{\Omega}\right) = \frac{\Lambda}{\Omega}, \label{eq:friedmann}
\eq
and relate the first three coefficients in the EFT action.
Generally, Eq.~(\ref{eq:eftaction}) introduces six time-dependent coefficients to which the FLRW metric Eq.~(\ref{eq:FLRW}) adds the scale factor $a(t)$, or the Hubble parameter $H(t)$.
Eqs.~(\ref{eq:friedmann}) introduce two constraints such that for specified matter content and spatial curvature the EFT formalism for linear Horndeski theory is composed of five free time-dependent functions~\cite{gubitosi:12,bloomfield:12,bellini:14,lombriser:14b}.
While one function describes the cosmological background evolution, the others specify the linear perturbations.
In order to describe the cosmological fluctuations implied by Eqs.~(\ref{eq:FLRW}) and (\ref{eq:eftaction}), the time diffeomorphism $t\rightarrow t+\pi(t,{\bf x})$ with scalar field perturbation $\pi$ is applied to the action to restore its full four-dimensional covariance.
The corresponding modified Einstein field equations can be found in Refs.~\cite{gubitosi:12,bloomfield:12}.
Importantly, the EFT coefficient $M_2^4(t)$ does not contribute if performing a quasistatic approximation in the fluctuations by neglecting time derivatives of the metric potentials and large-scale velocity flows with respect to spatial derivatives and matter density fluctuations, respectively~\cite{lombriser:14b}.

For computational convenience, we shall also employ a parametrisation developed in Ref.~\cite{bellini:14} that characterises the free time-dependent functions of the EFT action, Eq.~(\ref{eq:eftaction}), with more direct physical and observational implications.
We shall adopt the slightly different notation of Ref.~\cite{gleyzes:14b}.
The formalism separates out the expansion history $H$ as the free function determining the cosmological background, which is related to the EFT functions $\Omega$, $\Gamma$, and $\Lambda$ by Eqs.~(\ref{eq:friedmann}).
Linear perturbations around the background are then characterised by four additional free functions of time, $\alpha_i$, each implying a different physical effect in the properties of the scalar-tensor theory:
\bq
  \aK \equiv \frac{\Gamma+4M_2^4}{H^2 (\Omega+\Mbar_2^2)}, \ \ \ \aM \equiv \frac{\Omega' + (\Mbar_2^2)'}{\Omega + \Mbar_2^2}, \ \ \ \aB \equiv \frac{H\Omega' + \Mbar_1^3}{2H (\Omega+\Mbar_2^2)}, \ \ \ \aT \equiv -\frac{\Mbar_2^2}{\Omega+\Mbar_2^2}.
\label{eq:alphai}
\eq
The kineticity $\aK$ parametrises the contribution of a kinetic energy of the scalar field that causes the field to cluster at very large scales.
The evolution of the gravitational coupling with
the Planck mass
\bq
 M^2 \equiv \kappa^{-2}(\Omega + \Mbar_2^2)
\eq
at the rate $\aM$ gives rise
to a gravitational slip between the metric potentials $\Psi$ and $\Phi$,
attributing an effective anisotropic stress to the model and a modification in the damping of gravitational waves.
The braiding parameter $\aB$ captures the interaction of the scalar field and the metric through braiding, or mixing, of the kinetic contributions of these fields, causing the scalar field to cluster at small scales.
Finally, the tensor speed alteration $\aT$ parametrises the deviation of the speed of gravitational waves from the speed of light, which also acts as an effective anisotropic stress and causes a clustering of the scalar field.
As can be seen from Eq.~(\ref{eq:eftaction}), $\Lambda$CDM is recovered in the limit of $\alpha_i=0$ $\forall i$.
We refer to Refs.~\cite{bellini:14,gleyzes:14b,lombriser:15b} for the resulting perturbed modified Einstein equations in this formalism as well as a combination of the field equations into two modified Einstein equations and the energy-momentum conservation equations that eliminates contributions of $\pi$, its derivatives, and $\Psi'$.
This combination can be used to cast the modifications of the evolution of scalar modes in the form of Eqs.~(\ref{eq:mu}) through (\ref{eq:momcon}) after adopting a quasistatic or semi-dynamical approximation for $\Vm$ and the time derivatives of $\Phi$ (see Ref.~\cite{lombriser:15b} for more details).
From the EFT action, Eq.~(\ref{eq:eftaction}), and in the small-scale limit of $k\rightarrow\infty$, one obtains
\bqa
 \mu_{\infty} & = & \frac{2\left[\aB(1+\aT)-\aM+\aT\right]^2 + \alpha(1+\aT) c_{\rm s}^2}{\alpha \cs^2 \kappa^2M^2}, \label{eq:muinfH} \\
 \gamma_{\infty} & = & \frac{2\aB \left[\aB(1+\aT)-\aM+\aT\right] + \alpha c_{\rm s}^2}{2\left[\aB(1+\aT)-\aM+\aT\right]^2 + \alpha (1+\aT) \cs^2}, \label{eq:gammainfH}
\eqa
in Eqs.~(\ref{eq:mu}) and (\ref{eq:gamma}), respectively, where $\alpha \equiv 6 \aB^2 + \aK$ and the sound speed associated with the scalar field is
\bqa
 \cs^2 & = & -\frac{2}{\alpha} \left[ \vphantom{\frac{1'}{1}} \aB' + (1 + \aT)(1 + \aB)^2 - \left(1 + \aM - \frac{H'}{H}\right)(1 + \aB) + \frac{\rhom}{2H^2 M^2} \right]. \label{eq:cs}
\eqa
Here, $k\rightarrow\infty$ refers to a formal limit taken in linear theory, which in $\Lambda$CDM is typically applicable to scales of $k\lesssim0.1~\hMpc$.
The range of validity of linear computations can, however, vary with a modification of gravity~\cite{lombriser:13c,lombriser:15a}.
Note that contributions of $\alpha$ cancel in Eqs.~(\ref{eq:muinfH}) and (\ref{eq:gammainfH}) and hence $\aK$ does not affect the leading-order modifications at small scales.
At the next-to-leading order, velocity fields and time derivatives of $\Phi$ contribute such that a quasistatic approximation eventually fails but a semi-dynamical extension can be employed to account for corrections from these contributions~\cite{lombriser:15b}.
However, here we shall solve the full modified perturbed Einstein equations to derive the scalar fluctuations and use Eqs.~(\ref{eq:mu}) and (\ref{eq:gamma}) to determine $\mu(a,k)$ and $\gamma(a,k)$.
For the tensor fluctuations in Eq.~(\ref{eq:gw}) it follows that~\cite{gleyzes:14b,saltas:14}
\bq
 \nu=1+\aM, \ \ \ \ \ \ \cT^2=1+\aT,
\eq
which directly relates the modifications in the scalar modes implied by Horndeski gravity to a modification of the propagation of gravitational waves.
This relationship will become important to discriminate between a cosmological constant and a modification of gravity as the driver of cosmic acceleration in Sec.~\ref{sec:gw}.
Importantly, using Eqs.~(\ref{eq:alphai}) we can derive the relation
\bq
 \frac{\Omega'}{\Omega} = \aM + \frac{\aT'}{1+\aT}. \label{eq:nonzeroaMaT}
\eq
It implies that for a cosmic acceleration which is genuinely due to modified gravity, where $|\Omega'/\Omega|\gtrsim1$ at late times as described in Eqs.~(\ref{eq:accjordan}) and (\ref{eq:acceinstein}), at least one of either $\aM$ or $\aT'$ should be nonzero.

Finally note that the results presented here can easily be generalised to beyond-Horndeski theories~\cite{gleyzes:14a}, which introduce higher-order spatial derivatives in the modified Einstein equations, however, with a constraint equation ensuring a second-order equation for the propagating scalar degree of freedom.
In this case, the term $\delg\delta R^{(3)}$ in Eq.~(\ref{eq:eftaction}) acquires an extra EFT coefficient independent of $\Mbar_2^2(t)$, which can also be parametrised by the introduction of an additional parameter $\aH$ in Eqs.~(\ref{eq:alphai}), where $\aH=0$ represents the limit of Horndeski gravity~\cite{gleyzes:14b}.
Importantly, however, in beyond-Horndeski models the quasistatic approximation fails at leading order for $k\rightarrow\infty$ but Eqs.~(\ref{eq:muinfH}) and (\ref{eq:gammainfH}) can be generalised within a semi-dynamical description~\cite{lombriser:15b}.

\subsection{Nonlinear screening} \label{sec:nonlinear}

We have seen from Eqs.~(\ref{eq:alphai}) that our requirement of $\mathcal{O}(\Omega'/\Omega)\gtrsim1$ for self-acceleration implies that at least either $\aM$ or $\aT'$ are roughly of order unity at late times, and so we would naturally expect $\mathcal{O}(1)$ modifications in Eqs.~(\ref{eq:muinfH}) and (\ref{eq:gammainfH}).
Significant deviations from the GR values $\mu=\gamma=1$ can, however, not apply in the Solar-System region, where the gravitational dynamics has been well tested~\cite{will:05}.
In particular, a deviation from the GR value $\gamma_{\rm PPN}=1$ has been constrained at $\mathcal{O}(10^{-5})$ by the Shapiro time delay measured in the frequency shift of radio waves sent to and from the Cassini spacecraft~\cite{bertotti:03}.
Hence, the modification has to be screened in the nonlinear local region.
A variety of nonlinear screening mechanisms have been identified in scalar-tensor modifications of gravity.
They can be characterised by two main categories: a screening of the scalar field value such as is the case in chameleon~\cite{khoury:03a} models that acts through an appropriate form of the scalar field potential, or for symmetron~\cite{hinterbichler:10} and dilaton~\cite{brax:10} models with a particular form of the kinetic coupling in addition; the other screening mechanisms operate through derivative self-interactions of the scalar field such as is the case in Vainshtein screening~\cite{vainshtein:72} or k-mouflage models~\cite{babichev:09}.
The effects that the different screening mechanisms can have on nonlinear cosmic structure formation have been studied in detail in $N$-body simulations~\cite{winther:15,mead:14}, perturbation theory~\cite{koyama:09,brax:13}, as well as halo modelling~\cite{schmidt:08,schmidt:09b,lombriser:13b,lombriser:13c,barreira:14a} (see Ref.~\cite{lombriser:14a} for a review in chameleon gravity).

Models employing a screening mechanism operating through the suppression of the cosmological scalar field value in high-density regions have been found not to screen efficiently enough to accommodate both a self-acceleration and compatibility with Solar-System tests~\cite{hees:11,wang:12}.
At cosmological scales, the requirement of self-acceleration can furthermore introduce an anti-correlation between the integrated Sachs-Wolfe (ISW) effect and foreground galaxies~\cite{song:06,giannantonio:09,lombriser:10} that is inconsistent with measurements~\cite{ho:08,giannantonio:08}.
Note that for a wide range of chameleon gravity models cosmological constraints alone prohibit the modification as a genuine alternative explanation for cosmic acceleration to dark energy as $|\Omega'/\Omega|\ll1$~\cite{lombriser:14a}.
This independent conclusion is important as it implies that chameleon models may not serve as an alternative explanation for acceleration even if the chameleon field is not universally coupled, for instance, only coupling to dark matter and, hence, naturally satisfying the baryonic Solar-System constraints.

The Vainshtein mechanism has been broadly studied in the context of the Dvali-Gaba-dadze-Porrati (DGP)~\cite{dvali:00} braneworld model and in the generalisation of its effective four-dimensional scalar-tensor limit, the Galileon models~\cite{nicolis:08}, which both admit self-accelerating solutions.
The self-accelerated DGP model suffers from a ghost instability~\cite{koyama:07} and also introduces an observational tension between the cosmological background expansion history and the cosmic microwave background~\cite{fang:08}.
If allowing for the contribution of a cosmological constant or a brane tension $\Lambda$, observations clearly favour the constant over a five-dimensional bulk effect~\cite{lombriser:09}. 
Similarly, the quintic Galileon model has been found to provide an incomplete nonlinear solution~\cite{barreira:13}, the quartic model suppresses gravity in the high-density regime, yielding weak gravity that is incompatible with Solar-System tests~\cite{barreira:13}, and the cubic Galileon model is in observational tension with galaxy power spectra~\cite{barreira:14a} and produces a negative galaxy-ISW cross correlation~\cite{barreira:14b} inconsistent with observations.
It has furthermore been argued that all Horndeski and beyond-Horndeski models that are endowed with a Galilean shift-symmetry of the scalar field in Minkowski space, including the weakly broken galileons~\cite{pirtskhalava:15}, are strongly constrained by the Cassini measurement and the Hulse-Taylor pulsar due to modified gravitational dynamics that is not screened by the Vainshtein mechanism.

More generally, however, if the covariant theory is not specified a priori, the description of increasingly nonlinear structure requires an expansion of the EFT action, Eq.~(\ref{eq:eftaction}), to higher-order perturbations, each additional term, in principle, introducing a new free time-dependent coefficient.
It is therefore nontrivial to connect the nonlinear to the linear behaviour of general gravitational modifications for which the covariant action is not specified.
Moreover, for general Horndeski models, it can be shown that the screened and unscreened limits can be treated as independent gravitational theories as they are governed by different contributions in the gravitational action~\cite{mcmanus:15}.
Finally, Eq.~(\ref{eq:eftaction}) is in principle also more general than Horndeski theory as it may describe the background evolution and linear perturbations of a modified gravity model that is equivalent up to quadratic order in the action but differs in the nonlinear behaviour from Horndeski gravity.
Therefore, in order to place conservative constraints on the possibility of a self-acceleration through a scalar-tensor modification of gravity in general, we advocate the restriction to testing linear theory with linear observables, where both are well understood.

While a working nonlinear suppression mechanism may conservatively be assumed despite the challenges to the known mechanisms, possibly by involving both types of screening effects, the absence of evidence of modifications in the observed large-scale structure with increasingly precise measurements~\cite{planck:15} further questions the concept of cosmic acceleration from modified gravity.
Hence, featuring a nonlinear screening mechanism may potentially not be sufficient for a universally coupled and self-accelerated scalar-tensor theory to be observationally viable.

\subsection{Linear shielding mechanism} \label{sec:linshieldmech}

In Ref.~\cite{lombriser:14b} we have shown the existence of scalar-tensor theories that recover the $\Lambda$CDM values $\mu(a,k)=\gamma(a,k)=1$ in the quasistatic regime of linear perturbations.
They contain enough freedom to furthermore allow a matching of the concordance model background expansion history.
As discussed in Sec.~\ref{sec:eft}, the EFT formalism described by the action Eq.~(\ref{eq:eftaction}) and the FLRW metric Eq.~(\ref{eq:FLRW}) introduces five free functions of time, one of which is the scale factor $a(t)$, or equivalently the Hubble parameter $H(t)$.
In the following, we shall require that the scalar-tensor models reproduce the $\Lambda$CDM expansion history by fixing $H=H_{\Lambda{\rm CDM}}$.
Note that this implies an acceleration of $a(t)$ in the Jordan frame, Eq.~(\ref{eq:accjordan}), but not necessarily of $\tilde{a}(\tilde{t})$ in the Einstein--Friedmann frame, Eq.~(\ref{eq:acceinstein}), if $|\Omega'/\Omega|\gtrsim1$.
Importantly, the quasistatic limit only applies when time derivatives of the metric potentials can be neglected with respect to spatial derivatives and when matter density fluctuations dominate over large-scale velocity flows.
This generally applies at leading order to the linear cosmological perturbations of Horndeski theories in the small-scale limit but at the next-to-leading order, the velocity fields and time derivatives of the potentials can become important~\cite{lombriser:15b}.
We follow Ref.~\cite{lombriser:14b} and require that $\mu_{\infty}=\gamma_{\infty}=1$ to linearly recover $\Lambda$CDM in the formal limit of $k\rightarrow\infty$, which yields the linear shielding, or cancellation, conditions
\bqa
 \left(\Mbar_2^2\right)' & = & -\Omega' - \Mbar_2^2\left(1 + \frac{1}{2H}\frac{H\Omega'+\Mbar_1^3}{\Omega-1+\Mbar_2^2}\right), \label{eq:linshield1} \\
 \Gamma & = & H\left(1+\partial_{\ln a}\right)\Mbar_1^3 + 2H^2\left(\frac{H'}{H} - 1 - \partial_{\ln a} \right)\Mbar_2^2 \nonumber\\
 & & -\frac{ H\Omega' + \Mbar_1^3}{2(\Omega-1+\Mbar_2^2)}\left[H\Omega'-\Mbar_1^3+2H\left(1+\partial_{\ln a}\right)\Mbar_2^2\right]. \label{eq:linshield2}
\eqa
These relations characterise the $\mathcal{M}_{\rm II}$ class of linearly shielded scalar-tensor theories (see Ref.~\cite{lombriser:14b}) with the additional requirement of an embedding in Horndeski theory and $H=H_{\Lambda{\rm CDM}}$.
This reduces the EFT coefficients to two free functions of time of which one is $M_2^4(t)$ as it is not constrained by Eqs.~(\ref{eq:linshield1}) and (\ref{eq:linshield2}).
Note that the cancellation relations appear ad hoc in the EFT framework but in principle they may arise as a potential manifestation of a symmetry of a more fundamental theory.

While the linearly shielded models are degenerate with $\Lambda$CDM in the linear small-scale structure, deviations may still appear at ultra-large scales, where measurements are, however, typically limited by cosmic variance.
As has been proposed in Refs.~\cite{yoo:12,lombriser:13a}, information about gravitational interactions at these scales may be extracted from a multitracer analysis~\cite{seljak:08,mcdonald:08} that measures the relativistic effects in galaxy clustering~\cite{yoo:09,yoo:10,bonvin:11,challinor:11}.
Hereby galaxies in a galaxy-redshift survey are divided into differently biased samples, combining observations at different modes, and cross correlating the data with weak gravitational shear fields.
This procedure yields a measurement of the growth of matter density fluctuations that, in principle, is free of cosmic variance.
As has been demonstrated in Ref.~\cite{lombriser:13a}, such a measurement can provide new constraints on modified gravity and dark energy.
Deviations in the relativistic effects can be large in particular for models that intend to explain cosmic acceleration, where one would expect order unity modifications around the Hubble scale, as is the case, for instance, in the self-accelerated branch of DGP~\cite{hu:07b,fang:08,lombriser:09}.
This implies the computation beyond the quasistatic limit for which, more generally, one could employ the semi-dynamical approximation~\cite{lombriser:15b}.
However, note that for a model to cause any significant effects at these scales which can be competitively constrained compared to cosmological observations at smaller scales, and which are not already ruled out by current data, the modified gravity models need to employ some suppression mechanism like the linear cancellation effect discussed here.
Conversely, as we will discuss in Sec.~\ref{sec:model}, in some linearly shielded cases, not even a sample-variance free measurement of the ultra-large scale structure may be able to discriminate between a scalar-tensor modification of gravity and a cosmological constant as the driver of cosmic acceleration.

Instead of working with the EFT coefficients of Eq.~(\ref{eq:eftaction}), we may also adopt the four $\alpha_i$ functions of Eqs.~(\ref{eq:alphai}) with the fifth fixed to $H=H_{\Lambda{\rm CDM}}$.
The analogous relations to Eqs.~(\ref{eq:linshield1}) and (\ref{eq:linshield2}) can be derived from requiring $\mu_{\infty}=\gamma_{\infty}=1$ in Eqs.~(\ref{eq:muinfH}) and (\ref{eq:gammainfH}).
As $\aK$ does not affect the small-scale limit, it constitutes one of the two free functions of the resulting $\mathcal{M}_{\rm II}$ type models.
If adopting $\aB$ as the other, we obtain the linear shielding constraints
\bqa
 (M^2)' & = &  \aM M^2 = \aB\kappa^2M^4 - \frac{1-\kappa^2M^2}{\aB} \left\{ \vphantom{\left[\frac{1'}{1}\right]} \frac{\rhom}{2H^2} + \left[\aB' + \aB + (1+\aB)\frac{H'}{H}\right]M^2 \right\}, \label{eq:aM} \\
 \aT & = & \frac{\kappa^2M^2-1}{(1+\aB)\kappa^2M^2-1}\aM. \label{eq:aT}
\eqa
Note that if setting $\aB=0$, we only have the option of $M^2=1/\kappa^2$ with $\aM=\aT=0$.
Since $\aK$ is free, this scenario encompasses $\Lambda$CDM, quintessence, and $k$-essence models, although given the requirement of $H_{\Lambda{\rm CDM}}$ the quintessence models also reduce to $\Lambda$CDM.
There is a scenario in which $\aM=\aT=0$ but where $M^2\neq1/\kappa^2$, which leads to a differential equation for $\aB$.
However, as this implies a constant $\Omega$ from Eq.~(\ref{eq:nonzeroaMaT}), the model is not of interest to self-acceleration and we shall not pursue it further.
Importantly, note that for linear shielding to apply and $\Omega'\neq0$, we need non-vanishing $\aM$, $\aT$, and $\aB$.
In particular, this generally implies a deviation between the propagation speed of gravitational waves and the speed of light.

\section{A dark degeneracy} \label{sec:model}

As discussed in Sec.~\ref{sec:linshieldmech}, the EFT formalism of a linearly shielded Horndeski scalar-tensor theory that reproduces the concordance model background expansion history is defined by two free functions of time.
One of the functions necessarily is $M_2^4(t)$ or $\aK$ if using the parametrisation in Eqs.~(\ref{eq:alphai}).
We are interested in models for which the gravitational modification yields a genuine self-acceleration with $\mathcal{O}(\Omega'/\Omega)\gtrsim1$ that introduces a mapping from a non-accelerated Einstein--Friedmann frame scale factor $\rmd^2\tilde{a}/\rmd \tilde{t}^2\leq0$ to a Jordan frame acceleration $\rmd^2a/\rmd t^2>0$ in Eqs.~(\ref{eq:acceinstein}) and (\ref{eq:accjordan}).
Naturally, we therefore set the second free EFT coefficient by a parametrisation of $\Omega$.
We require that the modification only impacts the late universe and adopt a simple power law in terms of the scale factor $a(t)$ to describe the deviation from the GR value $\Omega=1$,
\bq
 \Omega = 1 + \Omega_+ a^n \label{eq:model}
\eq
with $n>0$.
Here, we fix $n=4$ such that the modification increases faster in time than $\rhom$ decreases.
The evolution of the Hubble parameter $H^2/H_0^2=\Om (a^{-3}-1)+1$ with $H_0=H(a=1)$ and $\kappa^2\rhom/3=H_0^2\Om a^{-3}$, together with the conformal mapping Eq.~(\ref{eq:model}), the Friedmann background evolution, Eqs.~(\ref{eq:friedmann}), determining $\Gamma(t)$ and $\Lambda(t)$, and the two linear shielding conditions~(\ref{eq:linshield1}) and (\ref{eq:linshield2}) fix four of the five EFT functions in Eq.~(\ref{eq:eftaction}).
The fifth EFT function $M_2^4(t)$ will be constrained by physicality conditions in Sec.~\ref{sec:properties}.
For illustration, we adopt the Planck value $\Om=0.308$~\cite{planck13:15} such that up to $M_2^4(t)$, the model is fully described by the parameter $\Omega_+$.

\subsection{Self-acceleration, linear cancellation, and stability conditions} \label{sec:properties}

\begin{figure}
 \centering
 \resizebox{0.5\hsize}{!}{\includegraphics{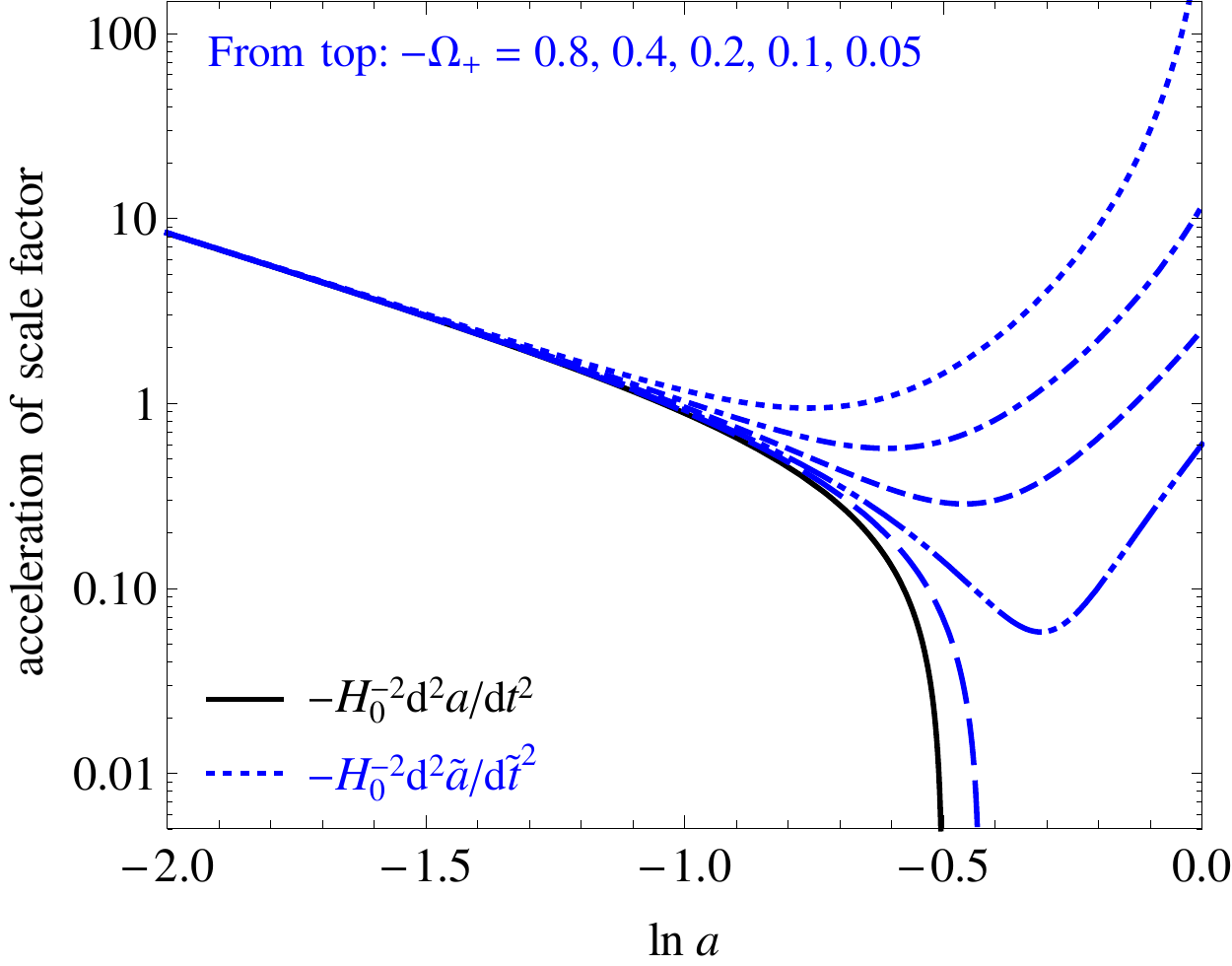}}
\caption{Cosmic acceleration from a modification of gravity.
In a genuinely self-accelerated modified gravity model, the conformal factor $\Omega$ maps a non-accelerated scale factor $\tilde{a}(\tilde{t})$ in the Einstein--Friedmann frame to the observed accelerated scale factor $a(t)$ in the Jordan frame, which matches here the expansion history $H(t)$ of the concordance cosmology $\Lambda$CDM.
For $\rmd^2\tilde{a}/\rmd\tilde{t}^2\lesssim0$ in the late-time universe, we require $\Omega_+\lesssim-0.1$.
\label{fig:selfacc}}
\end{figure}

Having specified $H$ as well as the conformal factor and EFT function $\Omega$, we evaluate $\rmd^2a/\rmd t^2$ and $\rmd^2\tilde{a}/\rmd \tilde{t}^2$ from Eqs.~(\ref{eq:accjordan}) and (\ref{eq:acceinstein}), respectively.
Since we adopted a $\Lambda$CDM evolution for $H$, we have an accelerated scale factor $a(t)$ in the Jordan frame by construction but the Einstein--Friedmann frame scale factor $\tilde{a}(\tilde{t})$ may not be accelerated depending on the magnitude of $\Omega_+$.
In Fig.~\ref{fig:selfacc}, we show the acceleration of the scale factor in the different frames as a function of $\ln a$ for varying values of $\Omega_+$.
From performing the conformal mapping, we find that for values of $\Omega_+\lesssim-0.1$, our Universe has not undergone a positive acceleration in the late-time cosmic history in the Einstein--Friedmann frame.
Hence, these models generate a sufficiently strong modification such that the observed acceleration may be attributed to a genuine modified gravity effect.
Note that this choice of $\Omega_+$ yields $|\Omega'/\Omega|\simeq0.5$ which is of the order of magnitude expected from the discussion in Sec.~\ref{sec:mg}.
Given $H/H_0$, $\Lambda/H_0^2$, and $\Gamma/H_0^2$, we solve the linear cancellation conditions in Eqs.~(\ref{eq:linshield1}) and (\ref{eq:linshield2}) for $\Mbar_1^3/H_0$ and $\Mbar_2^2/H_0^2$ with initial conditions set to recover GR at early times, $a_i=0.005$, where for simplicity we assume a matter-only universe.

\begin{figure*}
 \centering
 \resizebox{\hsize}{!}{
  \resizebox{0.98\hsize}{!}{\includegraphics{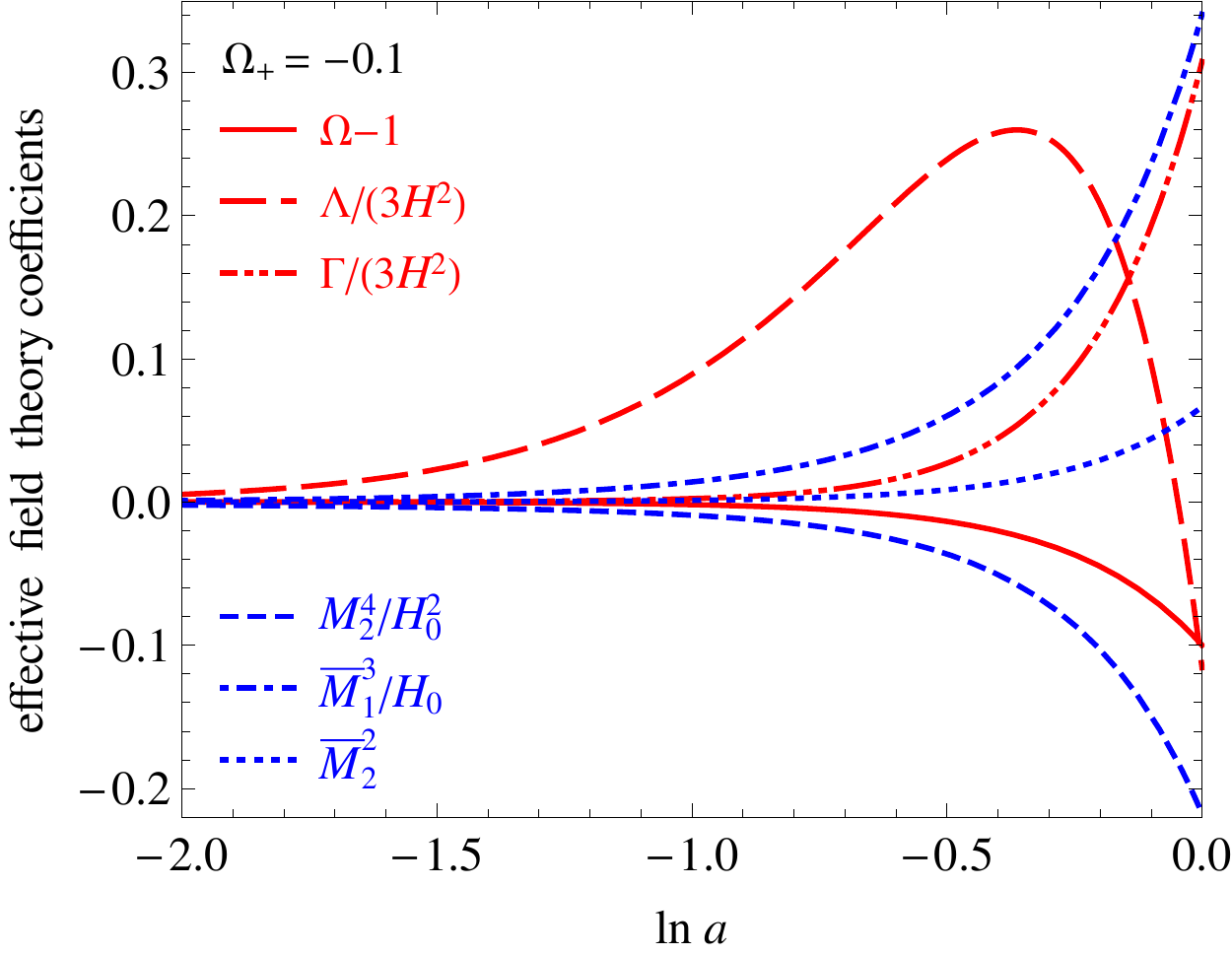}}
  \resizebox{\hsize}{!}{\includegraphics{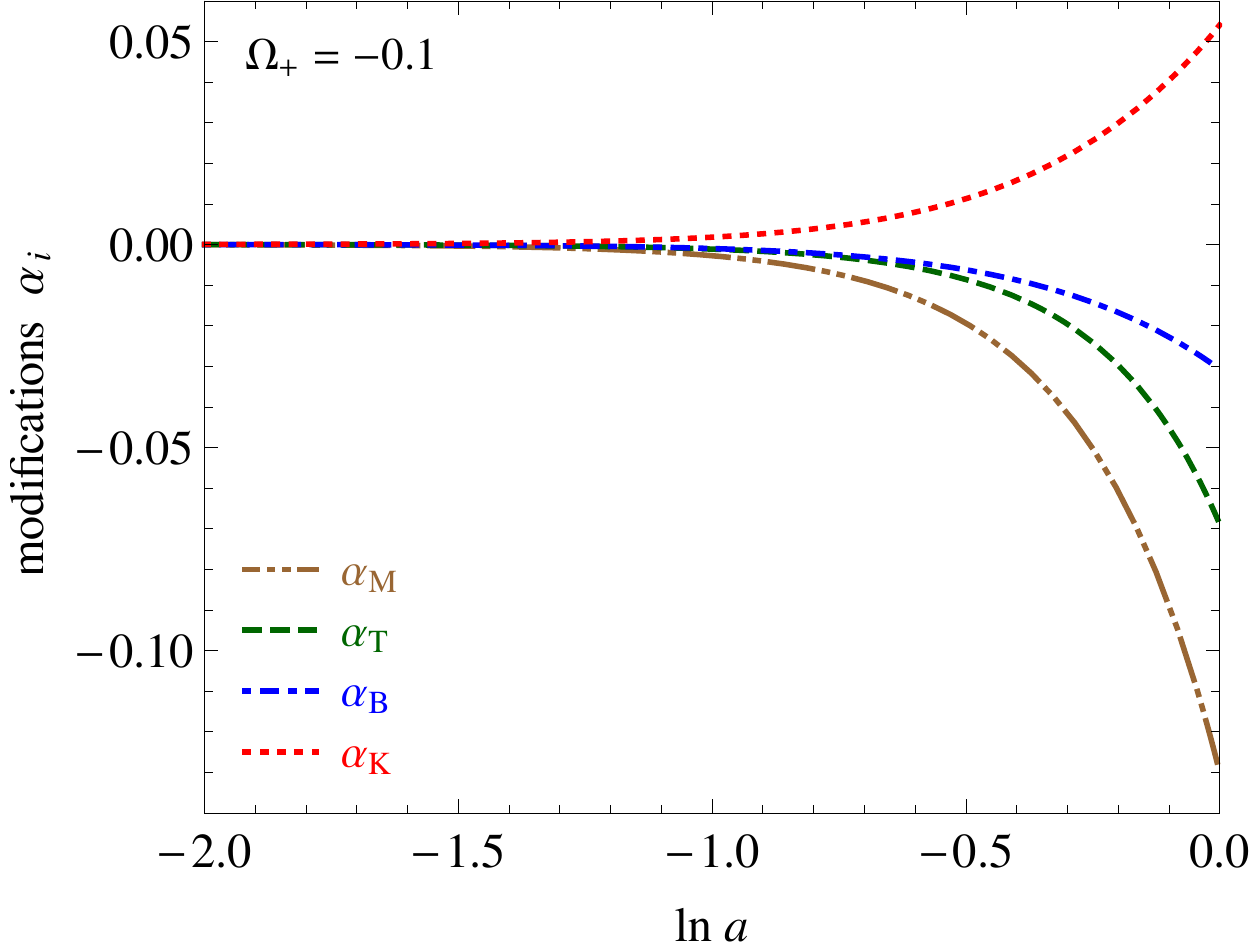}}
 }
\caption{The EFT coefficients (left panel) and modified gravity parameters $\alpha_i$ (right panel) characterising the cosmological background evolution and linear perturbations of the linearly shielded Horndeski scalar-tensor model.
The model recovers a GR cold dark matter universe at early times with deviations in the EFT coefficients from their $\Lambda$CDM values of order $\Omega_+$ at late times.
\label{fig:modifications}}
\end{figure*}

\begin{figure}
 \centering
 \resizebox{\hsize}{!}{
  \resizebox{\hsize}{!}{\includegraphics{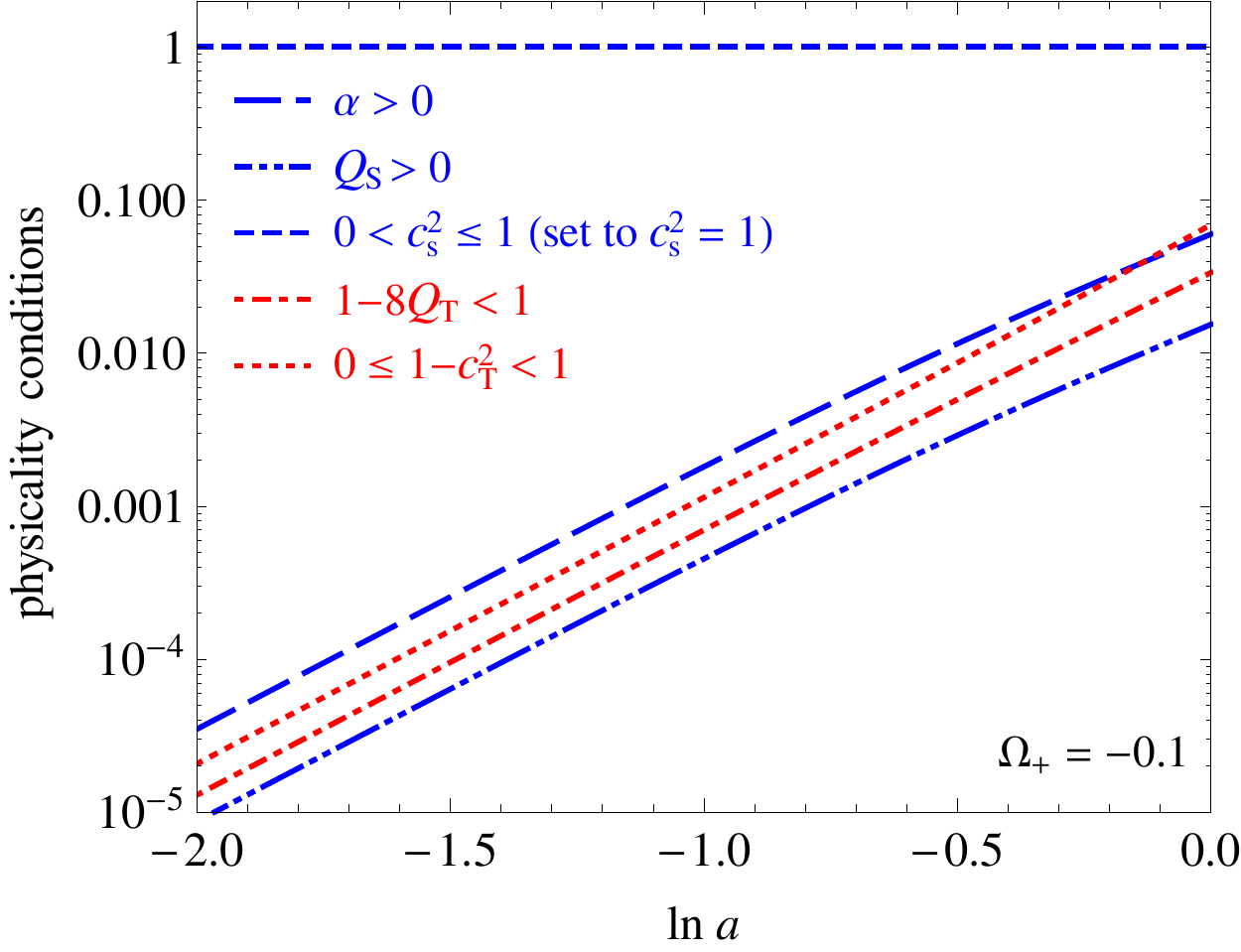}}
  \resizebox{0.988\hsize}{!}{\includegraphics{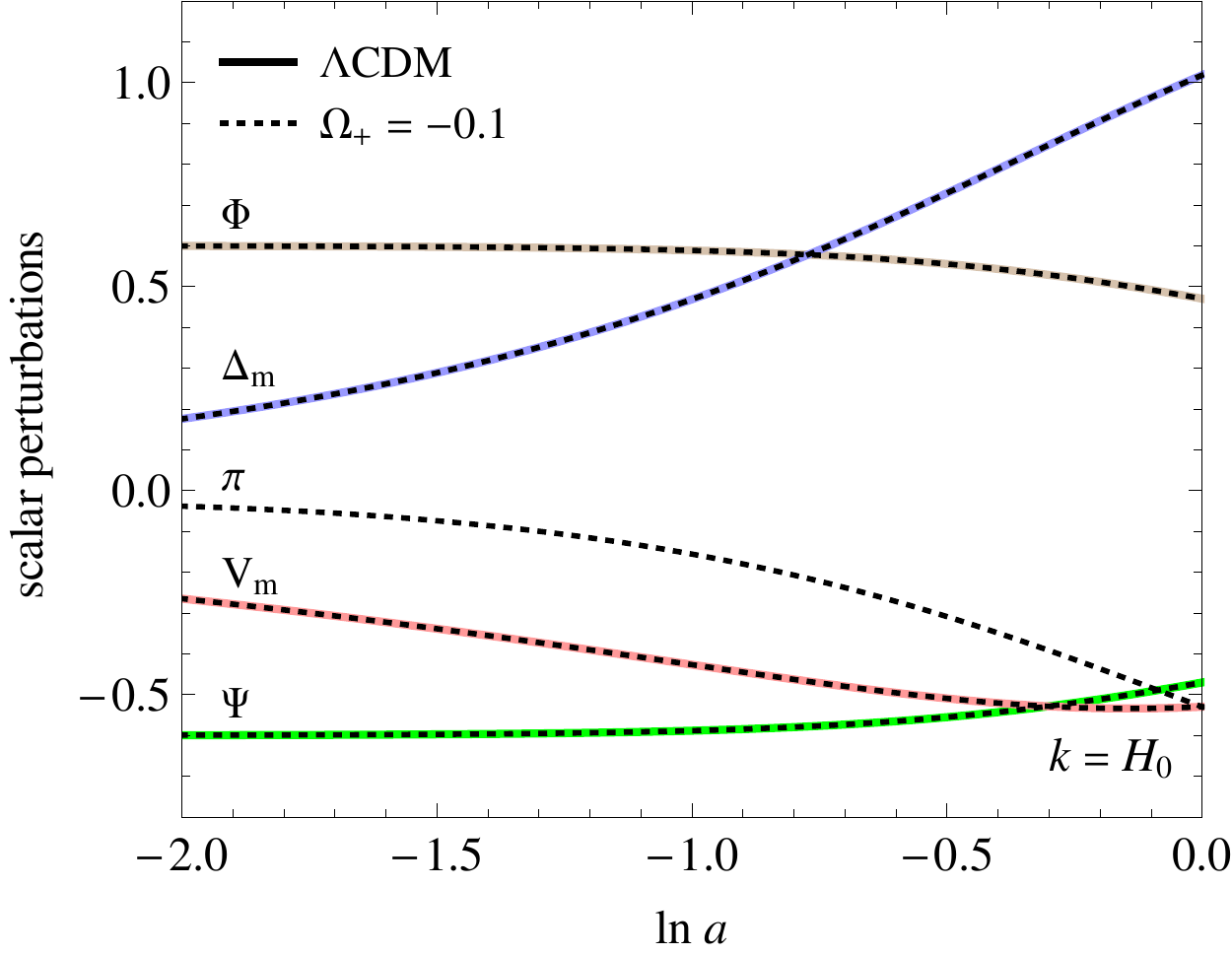}}
 }
\caption{\emph{Left panel:}
The linearly shielded scalar-tensor model satisfies stability conditions, whereby the sound speed of the scalar field fluctuation equals the speed of light $\cs^2=1$ and the tensor modes propagate at subluminal speed.
\emph{Right panel:} Comparison of scalar fluctuations in the linearly shielded model against $\Lambda$CDM.
For explicitness, we have evaluated the modified perturbation equations at the Hubble scale $k=H_0$, where one would typically expect signatures from a self-acceleration in the modified gravity model.
Deviations between the perturbative quantities $\epsilon$ in the two different models, however, remain within $|\epsilon/\epsilon_{\Lambda{\rm CDM}}-1|\lesssim2\times10^{-5}$ at all $a$.
This degeneracy extends to all scales with maximal deviations confined to the sub-percent level.
Hence, the model cannot be tested with ongoing and planned surveys of the large-scale structure.
\label{fig:physicality}}
\end{figure}

Next, we check that the background solution is stable against perturbations.
Generally, there may be ghost and gradient instabilities caused by a wrong sign of the kinetic contribution of the scalar field fluctuation and when its sound speed is imaginary, respectively.
To evade these instabilities in the scalar mode, we require that~\cite{bellini:14}
\bq
 2Q_{\rm S} \equiv \frac{M^2\alpha}{(1+\aB)^2}>0, \ \ \ \ \cs^2 > 0, \label{eq:scalstab}
\eq
where $\alpha$ and the sound speed $\cs$ have been defined in Sec.~\ref{sec:eft}.
To ensure the stability of the tensor modes, we need
\bq
 8 Q_{\rm T} \equiv M^2>0, \ \ \ \ \cT^2 = 1+\aT>0. \label{eq:tensstab}
\eq
Eqs.~(\ref{eq:scalstab}) and (\ref{eq:tensstab}) also imply that $\alpha>0$.
For the solutions to be physically well behaved, one may furthermore require that $\cs$ and $\cT$ should not be superluminal.
Note that a dependency on $\aK$ enters $\cs^2$ through $\alpha$ in Eq.~(\ref{eq:cs}) and hence on $M_2^4$ via Eqs.~(\ref{eq:alphai}).
In the following we shall require that the sound speed associated with the scalar field fluctuations equals the speed of light.
Thus, setting $\cs^2=1$ we can solve for $M_2^4$ and the linearly shielded modified gravity model becomes fully defined.
We show the resulting EFT coefficients and $\alpha_i$ parameters in Fig.~\ref{fig:modifications}.
The scalar-tensor model satisfies the stability constraints Eqs.~(\ref{eq:scalstab}) and (\ref{eq:tensstab}) as can be seen from the left-hand panel of Fig.~\ref{fig:physicality}.
At early times, it recovers GR.

\subsection{Degenerate large-scale structure} \label{sec:deglss}

As the model described in Sec.~\ref{sec:properties} satisfies the linear cancellation conditions in Eqs.~(\ref{eq:linshield1}) and (\ref{eq:linshield2}), the linear cosmological perturbations become indistinguishable from the concordance model in the small-scale regime, where $\mu_{\infty}=\gamma_{\infty}=1$.
Next, we focus on the modifications $\mu(a,k)$ and $\gamma(a,k)$ implied by the choice of EFT functions in Sec.~\ref{sec:properties} when $k$ becomes comparable to the Hubble scale and scales beyond it.
We first analyse the superhorizon limit $k\rightarrow0$ and solve the corresponding second-order homogeneous linear differential equation for $\Phi$ derived in Ref.~\cite{lombriser:15b} with the $\alpha_i$ parameters shown in Fig.~\ref{fig:modifications}, assuming a constant $\Phi_i$ set at an early time $a_i\ll1$ in the matter-dominated era.
We find that for the range of $\Omega_+$ values shown in Fig.~\ref{fig:selfacc}, $\Phi$ evolves equivalently to $\Lambda$CDM with deviations confined within the level of $\mathcal{O}(10^{-6})$ at any value of $a$.
However, importantly the coefficients of the differential equation for $\Phi$ in the scalar-tensor theory do not agree with the coefficients of the corresponding evolution equation for $\Phi$ in $\Lambda$CDM.
Hence, the degeneracy is encoded in the evolution of the EFT coefficients.
For the $\Lambda$CDM model the computed evolution applies to all linear scales, which is a consequence of momentum conservation and conservation of the comoving curvature fluctuation.

Next, we solve the full linearly perturbed modified Einstein equations implied by the EFT action, Eq.~(\ref{eq:eftaction}), given in the appendix of Ref.~\cite{lombriser:15b}, assuming GR initial conditions for the fluctuations at an early time $a_i\ll1$ in the matter-dominated era.
More specifically, we solve the energy-momentum conservation equations along with the time-time and the time derivative of the space-space components of the field equations, where we eliminate contributions of $\pi$ and $\pi'$ using the traceless space-space and time-space components, respectively~\cite{lombriser:15b}.
We evaluate the evolution of the perturbations for a range of wavenumbers spanning super- to sub-Hubble scales $k/H_0\in[0.1,100]$ in logarithmic spacings of $\Delta\log_{10}k=0.01$ and for the different values of $\Omega_+$ given in Fig.~\ref{fig:selfacc}.
At each $k$, we then compute $\mu(a,k)$ and $\gamma(a,k)$ by employing Eqs.~(\ref{eq:mu}) and (\ref{eq:gamma}).
We find no deviation from $\gamma=\mu=1$ exceeding the few permille level
throughout the cosmic history for any of the wavenumbers and any $\Omega_+$ values analysed.
Whether these deviations, oscillating around the $\Lambda$CDM values, are simply numerical artefacts or the degeneracy between the numerically evaluated perturbations of the models
can be shown to be mathematically exact remains subject to future work.
This may be an involved task, as in order to determine the fluctuations, we have employed a combination of modified perturbed Einstein equations, forming a coupled set of first-order linear differential equations whose coefficients are determined from solving the linear cancellation conditions that formulate a pair of coupled first-order nonlinear differential equations.
Note, however, that deviations of this small magnitude cannot be constrained by current cosmological data on the scalar fluctuations~\cite{planck:15} and for modified gravity models comparable to Eq.~(\ref{eq:model}), it has also been estimated that planned future surveys of the large-scale structure will only constrain $\mu$ and $\gamma$ at the level of 10\% and 30\%, respectively~\cite{song:10}.
Our results therefore imply the existence of a submodel of Horndeski scalar-tensor gravity, whose linear large-scale structure and background expansion history are indistinguishable from concordance cosmology, yet whose background evolution is self-accelerated.

The right-hand panel of Fig.~\ref{fig:physicality} shows the degeneracy between the linear cosmological perturbations of $\Lambda$CDM and the scalar-tensor models studied here, including a reconstruction of the scalar field fluctuation $\pi$ from the traceless space-space component.
For explicitness, we show the evolution of the modified gravity perturbations computed at the Hubble scale $k=H_0$, where one would typically expect large deviations for a self-accelerated modified gravity model but where departures from the $\Lambda$CDM perturbations remain within the level of $\lesssim2\times10^{-5}$.

\subsection{Nonlinear completion}

Finally, while the gravitational modifications in the model discussed here cancel at linear scales, this may no longer hold in the nonlinear regime since the scalar field itself is not screened, as can be seen in Fig.~\ref{fig:modifications}.
Linear cancellation models may therefore still need to invoke a nonlinear screening mechanism on nonlinear scales.
As discussed in Sec.~\ref{sec:nonlinear}, this is not an unfeasible requirement as an expansion of the EFT action to higher orders than second naturally introduces further free and independent coefficients, which then may be defined to yield a nonlinear shielding or cancellation effect.
A more natural approach to implement a nonlinear screening mechanism, however, would be to reconstruct a fully covariant scalar-tensor theory from the specified EFT coefficients in Fig.~\ref{fig:modifications}, for instance through re-introducing the scalar field $\phi$ by the replacement of $t\rightarrow\kappa^2\,\phi$ in the metric operators of Eq.~(\ref{eq:eftaction})~\cite{gubitosi:12}.
As the reconstruction of the nonlinear theory from the second-order action is not unique, the nonlinear freedom could then be used to implement a nonlinear screening effect.
However, the nonlinear completion of the linearly shielded model studied here lies beyond the scope of this paper and remains subject to future work.
As advocated in Sec.~\ref{sec:nonlinear} due to the ambiguity to connect nonlinear behaviour to linear theory and the accelerated expansion history, we conservatively focus on linear observables only.
Note that modified gravity models that would be shielded on linear scales but allow for some deviations in nonlinear observables before being screened again in the Solar System may provide an interesting phenomenology of its own even if not intended as explanation for cosmic acceleration, for instance, if considered in the context of $\Lambda$CDM problems in the substructure~\cite{lombriser:14c}.

\section{Breaking the degeneracy with gravitational waves} \label{sec:gw}

As emphasised in Sec.~\ref{sec:mg}, a modification of gravity may not just manifest itself in the cosmological large scale-structure but one should also expect a change in the propagation of tensor modes.
Therefore the measurement of gravitational waves emitted from events at cosmological distances may naturally be considered as a possibility to break the dark degeneracy in the large-scale structure and cosmic background expansion identified in Sec.~\ref{sec:model}.
More specifically, as we can see from Eq.~(\ref{eq:gw}), a running of the Planck mass $\aM$ impacts the damping term of the wave equation whereas $\aT$ changes the speed $\cT$ at which the gravitational wave propagates.
These effects have been used to constrain early modifications of gravity with the B-mode power spectrum of the cosmic microwave background~\cite{amendola:14,raveri:14,pettorino:14}.
However, as we focus on mechanisms for self-acceleration, we require either a direct detection of the primordial gravitational waves or constraints from other events at cosmological distances affected by late-time effects.
With the recent gravitational wave detection GW150914 from a merger of two black holes at $(0.4\pm0.2)$~Gpc with aLIGO~\cite{GW150914} such a constraint is obtained if the weak short gamma-ray burst measured by the Fermi GBM experiment~\cite{GRB150914} can be associated with the same event (GRB150914), requiring $\aT\simeq0$.
As we have seen in Sec.~\ref{sec:model}, self-acceleration of the degenerate model is intimately related to the changes in $\aM$ and $\aT$.
More specifically, it set a requirement on the modification of $\Omega_+\lesssim-0.1$, which implied $\alpha_{{\rm M}0}\lesssim-0.13$ and $\alpha_{{\rm T}0}\lesssim-0.07$ with $\alpha_{i0}=\alpha_i(a=1)$ as shown in Fig.~\ref{fig:modifications}.
Unlike the scalar fluctuations, however, the effects on the gravitational wave propagation are not degenerate, as we have checked by solving Eq.~(\ref{eq:gw}) with initial conditions at an early time $a_i\ll1$ in the matter-dominated regime.
Both frequency and amplitude of the tensor modes differ in $\Lambda$CDM.

In addition to the direct detection of gravitational waves, some indirect constraints can be inferred from astrophysical observations such as the orbital energy loss of binary pulsars that tests the coupling of matter sources to tensor waves.
However, as advocated in Sec.~\ref{sec:nonlinear}, for fully general constraints on the possibility of scalar-tensor modifications to explain cosmic acceleration, we conservatively assume that a screening mechanism may suppress the gravitational modifications in the galactic region and, hence, that the cosmological modification has no impact on the binary pulsars.
A constraint on the speed of gravitational waves can also be inferred from the observation of ultra-high energy cosmic rays of $\mathcal{O}(10^{11}~{\rm GeV})$.
If $\cT$ is subluminal, the cosmic rays should quickly lose their energy due to gravitational Cherenkov radiation at a rate of their energy squared and would not reach Earth.
If the cosmic rays are of galactic origin at around a 10~kpc distance, $\cT$ may only deviate at $\mathcal{O}(10^{-15})$ from the speed of light, and if of cosmological origin at around 2~Gpc, the constraint tightens to $\mathcal{O}(10^{-19})$~\cite{moore:01} (also see Ref.~\cite{caves:80}).
Note that these constraints only apply for subluminal $\cT$ as there would be no gravitational Cherenkov radiation otherwise.
Conservatively, we may not interpret these constraints as evidence against the presence of a cosmological scalar field with $\aT\neq0$ as in the case of galactic origin of these cosmic rays, the gravitational modifications could be screened such that the tensor modes propagate at the speed of light and no gravitational Cherenkov radiation is produced.
Moreover, as the constraints apply to gravitational waves with very short wavelengths of high energy they may not be directly applicable to the low-energy EFT of consideration here~\cite{jimenez:15}.

\subsection{Direct detection of gravitational waves} \label{sec:GWdetection}

With the recent gravitational wave detection GW150914 by aLIGO, we have entered the era of gravitational wave astronomy.
The radiation from astrophysical events will also be measured in other second-generation ground-based experiments such as the Advanced Virgo (aVIRGO)~\cite{aVIRGO} interferometer and the Kamioka Gravitational Wave Detector (KAGRA)~\cite{KAGRA} and then with the proposed third-generation space-based missions like the DECI-Hertz Interferometer Gravitational wave Observatory (DECIGO)~\cite{DECIGO} and the Evolved Laser Interferometer Space Antenna (eLISA)~\cite{eLISA}.
Further in the future the detectors may be improved with the Big Bang Observer (BBO)~\cite{BBO} and ground-based missions like LIGOIII~\cite{LIGOIII} and the Einstein Telescope (ET)~\cite{ET}.
Of particular interest are supernova events and short gamma-ray bursts for which both the gravitational wave and photon or neutrino emissions can be measured.
Second-generation observatories and current neutrino detectors operating over the next few years will measure the emissions from supernovae out to a distance of 100~kpc, roughly the size of the Milky Way.
Short gamma-ray bursts can be associated with binary mergers between neutron stars or between a neutron star and a black hole and with second-generation instruments the tensor waves emitted should be detectable within about 200~Mpc and 700~Mpc, respectively.
Note that the weak short gamma-ray burst measured by the Fermi GBM experiment simultaneously with GW150914 suggests that gamma-ray bursts may also be emitted by the merger of two black holes.
Supernovae need to be close enough for the gravitational waves to be measured and conversely, the gamma-ray bursts need to be aligned to allow the observation of the electromagnetic emission.
In both cases, the probability of a simultaneous detection of the different emissions is therefore low.
Within these volumes, one should only expect a few supernovae per century and a few simultaneous detections per century and per year of the tensor waves and electromagnetic radiation from the mergers between neutron stars or a merger with a black hole, respectively~\cite{nishizawa:14}, i.e., in addition to the possibly connected simultaneous observation of GW150914 from a black hole merger and GRB150914.
Future instruments may detect $\mathcal{O}(10^{5-6})$ of such binary mergers up to $z=2$ and $z=4$ or about 5~Gpc and 7~Gpc, respectively, mostly in the range of $z=1-3$, whereby one can anticipate hundreds to a thousand simultaneous observations of short gamma-ray bursts within a few years of operation~\cite{nishizawa:14,cutler:09}.

\subsection{Comparing arrival times} \label{sec:arrivaltimes}

With the simultaneous observation of the gravitational wave and electromagnetic or neutrino emissions from an astrophysical event, a difference in the speed of the tensor mode propagation from the speed of light as encountered in Sec.~\ref{sec:model} is naturally constrained by the direct comparison of the arrival times of the different emissions.
Importantly, although the number of simultaneous detections are expected to be low with the second-generation experiments described in Sec.~\ref{sec:GWdetection}, but likely with one already achieved by the GW+GRB150914 observations, we explain here how the confirmation of a single such observation suffices to break the dark degeneracy and as can be seen from Eq.~(\ref{eq:nonzeroaMaT}) rule out a large class of self-accelerated scalar-tensor theories.
To estimate the effect of the scalar-tensor modification on the relative arrival times, we assume that the binary mergers and the supernovae lie in well screened regions of space where GR is recovered and therefore deviations may only arise from the propagation of the gravitational wave through space.
There is an intrinsic delay between the emission times of the different radiative components but it is well understood from numerical studies and for the magnitude of the effect that we are concerned with here, where $\Delta\cT\equiv|\sqrt{1+\aT}-1|$ is in the few percent region, this delay is not relevant.
Likewise, detection timing errors, the mass effect of neutrinos, and distance uncertainties can also be neglected.
More specifically, for supernovae, the intrinsic time delay between the gravitational wave and neutrino emission is $\lesssim10^{-3}$~s, and a conservative estimate for the time delay between the tensor wave and photon emission in short gamma-ray bursts is $\lesssim500$~s~\cite{nishizawa:14}.  
While there may only be a few useful events per century from supernovae and a few per year from short gamma-ray bursts that can be measured with second-generation experiments as discussed in Sec.~\ref{sec:GWdetection}, a single simultaneous observation within the interval of the specific time delays places very tight constraints on $\Delta\cT$ of $\mathcal{O}(10^{-15})$ and $\mathcal{O}(10^{-14})$, respectively~\cite{nishizawa:14}.
These constraints apply to both subluminal and superluminal propagation and may recently have been achieved with the GW+GRB150914 observations.
Conservatively, the gravitational waves from supernovae measured with aLIGO or aVIRGO may be assumed to be screened since originating from within the Milky Way. 
The ET and next-generation neutrino detectors such as Hyper Kamioka Nucleon Decay Experiment (Hyper-Kamiokande)~\cite{abe:11} may reach 1~Mpc and improve the constraint on $\Delta\cT$ by about an order of magnitude.
However, this region may potentially still be screened by the local galaxy cluster such that the constraint may not be applicable in general.
For the cosmological distances measured with short gamma-ray bursts one also needs to take into account the cosmic expansion.
A dependence on redshift enters through the emission delays with the highest sensitivity reached around $1<z<3$ and for an event at $z=1$ or $\sim3$~Gpc, a constraint on $\Delta\cT$ at the level of $\mathcal{O}(10^{-15})$ can be inferred~\cite{nishizawa:14}.

Such constraints have profound implications for the degenerate models of Sec.~\ref{sec:model}.
For a measurement of the emission of a short gamma-ray burst at the cosmic distance of $z=1$, from the requirement of self-acceleration and with the redshift dependence of $\aT$, one would anticipate a $\gtrsim3\%$ effect on the propagation time which implies a delay in the arrival of the gravitational wave of $\gtrsim10^8$ years.
For a source at 200~Mpc, it would be a $\gtrsim0.5\%$ effect with a delay of more than several million years.
Hence, for the degenerate model discussed in Sec.~\ref{sec:model}, we cannot expect to measure the gravitational wave when observing the event in the electromagnetic emission.
Therefore, the association of the gravitational wave GW150914 from a merger of two black holes at $\sim400$~Mpc with the weak short gamma-ray burst GRB150904, if confirmed, may be the crucial observation that breaks the dark degeneracy.

\subsection{Standard sirens} \label{sec:standardsirens}

Besides the relative arrival time of the cosmological gravitational wave signal to its electromagnetic counterpart, the measurement of the amplitude and frequency of the tensor wave as well as their evolution allows to infer further constraints on cosmology, dark energy, and gravity.
The chirping time of an inspiral or merger event together with its orbital frequency and strain provides a measurement of the luminosity distance $\DL$ to the source from the decay of the gravitational wave amplitude with $1/\DL$.
If combined with a measurement of the redshift of the source and with enough samples, this can yield a high-precision probe of the cosmological background expansion and serve as a standard siren~\cite{schutz:86,holz:05}.
The difficulty, however, lies in identifying the host galaxies of the binaries to associate the correct redshifts to the distance measure.
The electromagnetic emission of short-gamma ray bursts mentioned in Sec.~\ref{sec:GWdetection} could be used for this purpose. 
The second-generation detectors may only provide a dozen such samples that will not suffice for a precision measurement of $\DL(z)$.
In future experiments for about a thousand of them a redshift may be inferred and with decreasing detection error in the volume, an identification of the host galaxy may also become possible without electromagnetic counterpart from the event and increase the number of standard sirens to $\mathcal{O}(10^{5-6})$, enabling precision tests of cosmology.
An identification of the redshifts of the gravitational wave signals without electromagnetic measurement may also be possible through the Doppler-modulation caused by the motion of the instrument around the Sun.
The eLISA should detect dozens of inspirals of supermassive black holes per year and may use the effect to make about one such observation every year for objects within $z\lesssim1$~\cite{eLISA}.
Cosmological tests with standard sirens are limited by the magnification effect from gravitational lensing generated by the matter distribution between the source and the observer.
For the second-generation detectors the effect is less important as their detection range is limited to about 1~Gpc but for future experiments observing sources at $\gtrsim3$~Gpc, the effect on $\DL$ can reach a few percent.
The magnification can be modelled from weak lensing maps generated by measurements of the deformation of galaxy images.
With high enough angular resolution of the gravitational wave detector, the effect could also be used as an additional probe of the growth of cosmic structure.
Second-generation experiments should allow a measurement of the luminosity distance $\Delta\DL/\DL$ within several percent while future instruments should achieve percent-level precision with the error dominated by the magnification effect~\cite{cutler:09,shapiro:09}.

While providing a precision probe for the cosmological background evolution, standard sirens will also test potential modifications of gravity.
As discussed in Sec.~\ref{sec:mg}, a running of the Planck mass $\aM\neq0$ introduces a deviation in the damping term of the gravitational wave equation, Eq.~(\ref{eq:gw}), which affects the amplitude of the observed tensor wave, and together with $\aT\neq0$ also changes the observed frequency of the wave.
We shall assume that the mergers lie in a well-screened regime where GR is recovered and that hence the gravitational modification only impacts the propagation of the gravitational wave through space as described by Eq.~(\ref{eq:gw}).
For simplicity, we also assume that the decay in amplitude can be measured at the same precision as in GR.
In order to estimate the effect of nonzero $\aM$ on this decay, we shall further approximate the damping term and its modification as constant effects averaged over the redshift to the source.
For the degenerate model of Sec.~\ref{sec:model} with $\alpha_{{\rm M}0}\lesssim-0.13$ and for a source at $z=1$, the decay of the gravitational wave amplitude becomes $\gtrsim5\%$ less efficient.
With the precision of $\Delta\DL/\DL\approx1\%$ that can be inferred from the decay of the amplitude in GR in future experiments, we likewise estimate that the gravitational modification would roughly be detectable at the $5\sigma$-level, provided that the background parameters can be sufficiently constrained by complementary electromagnetic measurements.
Note that the dominating magnification error of the measurement is not affected by the modification due to the degeneracy in the large-scale structure.

\subsection{Implications for self-accelerated scalar-tensor gravity}

\begin{figure}
 \centering
 \resizebox{0.55\hsize}{!}{\includegraphics{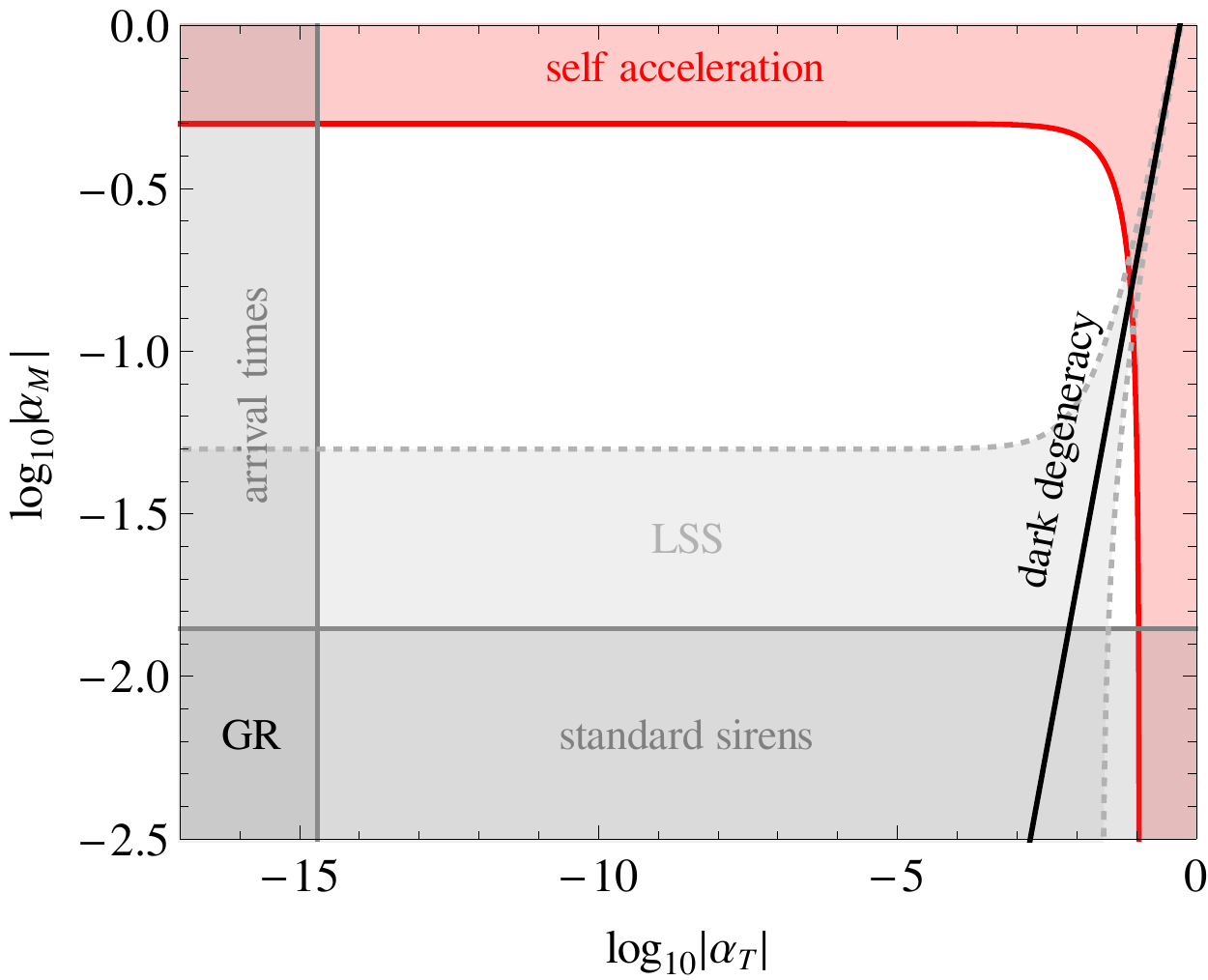}}
\caption{Prospective constraints on a running Planck mass $\aM\leq0$ and a modified speed of tensor mode propagation $-1<\aT\leq0$ with gravitational wave detectors measuring the decay of the wave amplitude with standard sirens and the arrival time with respect to the electromagnetic emission from a cosmological event.
Shaded regions indicate allowed regions.
As described in Eq.~(\ref{eq:nonzeroaMaT}) scalar-tensor theories require at least either $\aM$ or $\aT'$ to be $\mathcal{O}(0.1-1)$ for a self-accelerated cosmic expansion due to modified gravity, here schematically set by $|\Omega'/\Omega|_{z=0}\gtrsim0.5$ and assuming $\aT'/\aT\approx4$.
The dark degeneracy model introduced in Sec.~\ref{sec:model} cannot be discriminated from $\Lambda$CDM with measurements of the cosmological background expansion and large-scale structure but its self-accelerating regime can be tested with gravitational wave cosmology.
More generally, the simultaneous measurement of the gravitational wave and electromagnetic emissions of a cosmological event implies $\aT\simeq0$ and limits self-acceleration to an effect due to $\mathcal{O}(\aM)\gtrsim0.5$, which could either be ruled out by future constraints from large-scale structure (LSS) or from standard sirens, challenging the concept of self-acceleration from a scalar-tensor modification of gravity.
Note that such a simultaneous measurement may potentially already have been achieved with the recent gravitational wave observation GW150914 with aLIGO~\cite{GW150914} and a weak short gamma-ray burst measured by the Fermi GBM experiment~\cite{GRB150914}.
\label{fig:GWconstraints}}
\end{figure}

A measurement of both $\aM$ and $\aT$ from standard sirens and the arrival times of the gravitational wave and electromagnetic signals also has more generally important implications for the concept of self-acceleration through a scalar-tensor modification of gravity.
For instance, a simultaneous arrival of the different emissions not only strongly constrains the available EFT model space by requiring $\aT\simeq0$ but also limits the self-accelerated scalar-tensor theories to models with $\mathcal{O}(\aM)\gtrsim(0.1-1)$ at late times as can be seen from Eq.~(\ref{eq:nonzeroaMaT}).
Such models may be constrained by standard sirens but as they must also show a signature in the large-scale structure as discussed in Sec.~\ref{sec:linshieldmech}, for which strong independent electromagnetic constraints will be available, a single measurement of a gravitational wave from a cosmological event with electromagnetic counterpart may likely rule out the idea that cosmic acceleration could be due to a gravitational modification from scalar-tensor gravity, at least for the Horndeski models considered here.
The GW+GRB150914 detections may likely have provided this crucial measurement.
As pointed out in Sec.~\ref{sec:deglss}, future surveys of the large-scale structure will place constraints on modifications of the scalar fluctuations comparable to Eq.~(\ref{eq:model}) that are of the order of 10\% in $\mu$ and 30\% in $\gamma$~\cite{song:10}.
This approximately translates to constraints on $\aM$ and $\aT$ that are at the level of 5\%.
We schematically illustrate the situation in Fig.~\ref{fig:GWconstraints}.
Note that Eq.~(\ref{eq:nonzeroaMaT}) also holds in beyond-Horndeski models and a similar conclusion applies unless $\aH$ can cancel modifications in the large-scale structure and limit constraints on $\aM$ to tests with standard sirens.

\section{Conclusions} \label{sec:conclusions}

We have identified a Horndeski scalar-tensor model whose cosmological background expansion and linear scalar perturbations are indistinguishable from the concordance cosmology.
The model can be characterised by a single new parameter that describes the magnitude of the conformal mapping between the Einstein--Friedmann and Jordan frames of the background cosmology.
If this parameter exceeds a specified threshold, the model provides a late-time acceleration of the cosmic expansion that can genuinely be attributed to a modification of gravity rather than a cosmological constant or dark energy as it does not show any acceleration in the Einstein--Friedmann frame.
The linear model is furthermore free of ghost and gradient instabilities and attributes a sound speed to the scalar field fluctuations which equals the speed of light.
While degenerate in the scalar modes, the requirement of self-acceleration implies a present propagation speed of gravitational waves emitted by cosmological sources at $\lesssim95\%$ of the speed of light.
It also predicts a damping of the wave amplitude that is $\gtrsim5\%$ less efficient than in GR.
We have discussed the observability of these effects with current and future gravitational wave detectors and the implication of a potential association of the recent gravitational wave detection GW150914 by aLIGO with a weak short gamma-ray burst measured by the Fermi GBM experiment.

Our results suggest that the surveys mapping the cosmological structure of our Universe and measurements of the cosmic background evolution alone are not sufficient to ultimately discriminate between a cosmological constant or a dark energy and a modification of gravity as the driver of cosmic acceleration.
In order to draw general and conservative conclusions, we have limited our discussion to linear theory as modifications at nonlinear scales may be affected by complex screening effects, potentially involving different shielding mechanisms in different astrophysical regimes, which limits a general connection to the modifications driving the cosmological background evolution.
Whether the dark degeneracy can generally be broken by the consideration of nonlinear observables remains subject to future work.
Here, we have shown that at least for Horndeski scalar-tensor theories this linear degeneracy in the explanation of cosmic acceleration can be broken by a single simultaneous measurement of the gravitational wave and electromagnetic emissions from an event at a cosmological distance, likely recently achieved by the simultaneous GW+GRB150914 observations. 
We have furthermore described how such a measurement may more generally imply a severe challenge to the concept of a genuine self-acceleration due to a scalar-tensor modification of gravity.
With second-generation gravitational wave detectors in operation, if not already succeeded, it is conceivable that such a measurement may be done within the next few years.

\section*{Acknowledgement}

This work was supported by the U.K.~STFC Consolidated Grant for Astronomy and Astrophysics at the University of Edinburgh.
LL acknowledges the support of a SNSF Advanced Postdoc.Mobility Fellowship (No.~161058).
AT acknowledges the support of a Royal Society Wolfson Research Merit Award.
Please contact the authors for access to research materials.

\bibliographystyle{JHEP}
\bibliography{selfaccdeg}

\providecommand{\href}[2]{#2}\begingroup\raggedright\begin{thebibliography}{10}

\bibitem{will:05}
C.~M. Will, {\it {The confrontation between general relativity and
  experiment}},  {\em Living Rev. Rel.} {\bf 9} (2005) 3,
  [\href{http://arxiv.org/abs/gr-qc/0510072}{{\tt gr-qc/0510072}}].

\bibitem{vainshtein:72}
A.~Vainshtein, {\it {To the problem of nonvanishing gravitation mass}},  {\em
  Phys. Lett.} {\bf B39} (1972) 393--394.

\bibitem{khoury:03a}
J.~Khoury and A.~Weltman, {\it {Chameleon fields: Awaiting surprises for tests
  of gravity in space}},  {\em Phys. Rev. Lett.} {\bf 93} (2004) 171104,
  [\href{http://arxiv.org/abs/astro-ph/0309300}{{\tt astro-ph/0309300}}].

\bibitem{babichev:09}
E.~{Babichev}, C.~{Deffayet}, and R.~{Ziour}, {\it {k-MOUFLAGE Gravity}},  {\em
  IJMPD} {\bf 18} (2009) 2147--2154,
  [\href{http://arxiv.org/abs/0905.2943}{{\tt arXiv:0905.2943}}].

\bibitem{hinterbichler:10}
K.~Hinterbichler and J.~Khoury, {\it {Symmetron Fields: Screening Long-Range
  Forces Through Local Symmetry Restoration}},  {\em Phys. Rev. Lett.} {\bf
  104} (2010) 231301, [\href{http://arxiv.org/abs/1001.4525}{{\tt
  arXiv:1001.4525}}].

\bibitem{brax:10}
P.~Brax, C.~van~de Bruck, A.-C. Davis, and D.~Shaw, {\it {The Dilaton and
  Modified Gravity}},  {\em Phys. Rev.} {\bf D82} (2010) 063519,
  [\href{http://arxiv.org/abs/1005.3735}{{\tt arXiv:1005.3735}}].

\bibitem{lombriser:14b}
L.~Lombriser and A.~Taylor, {\it {Classifying Linearly Shielded Modified
  Gravity Models in Effective Field Theory}},  {\em Phys.Rev.Lett.} {\bf 114}
  (2015), no.~3 031101, [\href{http://arxiv.org/abs/1405.2896}{{\tt
  arXiv:1405.2896}}].

\bibitem{hees:11}
A.~Hees and A.~Fuzfa, {\it {Combined cosmological and solar system constraints
  on chameleon mechanism}},  {\em Phys. Rev.} {\bf D85} (2012) 103005,
  [\href{http://arxiv.org/abs/1111.4784}{{\tt arXiv:1111.4784}}].

\bibitem{wang:12}
J.~Wang, L.~Hui, and J.~Khoury, {\it {No-Go Theorems for Generalized Chameleon
  Field Theories}},  {\em Phys. Rev. Lett.} {\bf 109} (2012) 241301,
  [\href{http://arxiv.org/abs/1208.4612}{{\tt arXiv:1208.4612}}].

\bibitem{dvali:00}
G.~Dvali, G.~Gabadadze, and M.~Porrati, {\it {4-D gravity on a brane in 5-D
  Minkowski space}},  {\em Phys.Lett.} {\bf B485} (2000) 208--214,
  [\href{http://arxiv.org/abs/hep-th/0005016}{{\tt hep-th/0005016}}].

\bibitem{deffayet:00}
C.~Deffayet, {\it {Cosmology on a brane in Minkowski bulk}},  {\em Phys. Lett.}
  {\bf B502} (2001) 199--208, [\href{http://arxiv.org/abs/hep-th/0010186}{{\tt
  hep-th/0010186}}].

\bibitem{nicolis:08}
A.~Nicolis, R.~Rattazzi, and E.~Trincherini, {\it {The Galileon as a local
  modification of gravity}},  {\em Phys. Rev.} {\bf D79} (2009) 064036,
  [\href{http://arxiv.org/abs/0811.2197}{{\tt arXiv:0811.2197}}].

\bibitem{pirtskhalava:15}
D.~Pirtskhalava, L.~Santoni, E.~Trincherini, and F.~Vernizzi, {\it {Weakly
  Broken Galileon Symmetry}},  {\em JCAP} {\bf 1509} (2015), no.~09 007,
  [\href{http://arxiv.org/abs/1505.00007}{{\tt arXiv:1505.00007}}].

\bibitem{koyama:07}
K.~Koyama, {\it {Ghosts in the self-accelerating universe}},  {\em Class.
  Quant. Grav.} {\bf 24} (2007) R231--R253,
  [\href{http://arxiv.org/abs/0709.2399}{{\tt arXiv:0709.2399}}].

\bibitem{barreira:13}
A.~Barreira, B.~Li, C.~M. Baugh, and S.~Pascoli, {\it {Spherical collapse in
  Galileon gravity: fifth force solutions, halo mass function and halo bias}},
  {\em JCAP} {\bf 1311} (2013) 056, [\href{http://arxiv.org/abs/1308.3699}{{\tt
  arXiv:1308.3699}}].

\bibitem{jimenez:15}
J.~B. Jim\'enez, F.~Piazza, and H.~Velten, {\it {Piercing the Vainshtein screen
  with anomalous gravitational wave speed: Constraints on modified gravity from
  binary pulsars}},  {\em Phys. Rev. Lett.} {\bf 116} (2016) 061101,
  [\href{http://arxiv.org/abs/1507.05047}{{\tt arXiv:1507.05047}}].

\bibitem{fang:08}
W.~Fang, S.~Wang, W.~Hu, Z.~Haiman, L.~Hui, and M.~May, {\it {Challenges to the
  DGP Model from Horizon-Scale Growth and Geometry}},  {\em Phys. Rev.} {\bf
  D78} (2008) 103509, [\href{http://arxiv.org/abs/0808.2208}{{\tt
  arXiv:0808.2208}}].

\bibitem{lombriser:09}
L.~Lombriser, W.~Hu, W.~Fang, and U.~Seljak, {\it {Cosmological Constraints on
  DGP Braneworld Gravity with Brane Tension}},  {\em Phys. Rev.} {\bf D80}
  (2009) 063536, [\href{http://arxiv.org/abs/0905.1112}{{\tt
  arXiv:0905.1112}}].

\bibitem{barreira:14a}
A.~Barreira, B.~Li, W.~A. Hellwing, L.~Lombriser, C.~M. Baugh, and S.~Pascoli,
  {\it {Halo model and halo properties in Galileon gravity cosmologies}},  {\em
  JCAP} {\bf 1404} (2014) 029, [\href{http://arxiv.org/abs/1401.1497}{{\tt
  arXiv:1401.1497}}].

\bibitem{planck:15}
{\bf Planck} Collaboration, P.~Ade et~al., {\it {Planck 2015 results. XIV. Dark
  energy and modified gravity}},  \href{http://arxiv.org/abs/1502.01590}{{\tt
  arXiv:1502.01590}}.

\bibitem{ratra:87}
B.~Ratra and P.~J.~E. Peebles, {\it {Cosmological Consequences of a Rolling
  Homogeneous Scalar Field}},  {\em Phys. Rev.} {\bf D37} (1988) 3406.

\bibitem{wetterich:87}
C.~Wetterich, {\it {Cosmology and the Fate of Dilatation Symmetry}},  {\em
  Nucl. Phys.} {\bf B302} (1988) 668.

\bibitem{armendariz:00}
C.~Armendariz-Picon, V.~F. Mukhanov, and P.~J. Steinhardt, {\it {Essentials of
  k essence}},  {\em Phys. Rev.} {\bf D63} (2001) 103510,
  [\href{http://arxiv.org/abs/astro-ph/0006373}{{\tt astro-ph/0006373}}].

\bibitem{creminelli:08}
P.~Creminelli, G.~D'Amico, J.~Norena, and F.~Vernizzi, {\it {The Effective
  Theory of Quintessence: the $w<-1$ Side Unveiled}},  {\em JCAP} {\bf 0902}
  (2009) 018, [\href{http://arxiv.org/abs/0811.0827}{{\tt arXiv:0811.0827}}].

\bibitem{park:10}
M.~Park, K.~M. Zurek, and S.~Watson, {\it {A Unified Approach to Cosmic
  Acceleration}},  {\em Phys.Rev.} {\bf D81} (2010) 124008,
  [\href{http://arxiv.org/abs/1003.1722}{{\tt arXiv:1003.1722}}].

\bibitem{gubitosi:12}
G.~Gubitosi, F.~Piazza, and F.~Vernizzi, {\it {The Effective Field Theory of
  Dark Energy}},  {\em JCAP} {\bf 1302} (2013) 032,
  [\href{http://arxiv.org/abs/1210.0201}{{\tt arXiv:1210.0201}}].

\bibitem{bloomfield:12}
J.~K. Bloomfield, E.~E. Flanagan, M.~Park, and S.~Watson, {\it {Dark energy or
  modified gravity? An effective field theory approach}},  {\em JCAP} {\bf
  1308} (2013) 010, [\href{http://arxiv.org/abs/1211.7054}{{\tt
  arXiv:1211.7054}}].

\bibitem{bellini:14}
E.~Bellini and I.~Sawicki, {\it {Maximal freedom at minimum cost: linear
  large-scale structure in general modifications of gravity}},  {\em JCAP} {\bf
  1407} (2014) 050, [\href{http://arxiv.org/abs/1404.3713}{{\tt
  arXiv:1404.3713}}].

\bibitem{gleyzes:14b}
J.~Gleyzes, D.~Langlois, and F.~Vernizzi, {\it {A unifying description of dark
  energy}},  {\em Int.J.Mod.Phys.} {\bf D23} (2014) 3010,
  [\href{http://arxiv.org/abs/1411.3712}{{\tt arXiv:1411.3712}}].

\bibitem{horndeski:74}
G.~W. Horndeski, {\it {Second-order scalar-tensor field equations in a
  four-dimensional space}},  {\em Int.J.Theor.Phys.} {\bf 10} (1974) 363--384.

\bibitem{deffayet:11}
C.~Deffayet, X.~Gao, D.~Steer, and G.~Zahariade, {\it {From k-essence to
  generalised Galileons}},  {\em Phys.Rev.} {\bf D84} (2011) 064039,
  [\href{http://arxiv.org/abs/1103.3260}{{\tt arXiv:1103.3260}}].

\bibitem{kobayashi:11}
T.~Kobayashi, M.~Yamaguchi, and J.~Yokoyama, {\it {Generalized G-inflation:
  Inflation with the most general second-order field equations}},  {\em Prog.
  Theor. Phys.} {\bf 126} (2011) 511--529,
  [\href{http://arxiv.org/abs/1105.5723}{{\tt arXiv:1105.5723}}].

\bibitem{GW150914}
{\bf Virgo, LIGO Scientific} Collaboration, B.~Abbott et~al., {\it {Observation
  of Gravitational Waves from a Binary Black Hole Merger}},  {\em Phys. Rev.
  Lett.} {\bf 116} (2016), no.~6 061102,
  [\href{http://arxiv.org/abs/1602.03837}{{\tt arXiv:1602.03837}}].

\bibitem{GRB150914}
V.~Connaughton et~al., {\it {Fermi GBM Observations of LIGO Gravitational Wave
  event GW150914}},  \href{http://arxiv.org/abs/1602.03920}{{\tt
  arXiv:1602.03920}}.

\bibitem{kunz:07}
M.~Kunz, {\it {The dark degeneracy: On the number and nature of dark
  components}},  {\em Phys. Rev.} {\bf D80} (2009) 123001,
  [\href{http://arxiv.org/abs/astro-ph/0702615}{{\tt astro-ph/0702615}}].

\bibitem{bettoni:13}
D.~Bettoni and S.~Liberati, {\it {Disformal invariance of second order
  scalar-tensor theories: Framing the Horndeski action}},  {\em Phys. Rev.}
  {\bf D88} (2013) 084020, [\href{http://arxiv.org/abs/1306.6724}{{\tt
  arXiv:1306.6724}}].

\bibitem{uzan:06}
J.-P. Uzan, {\it {The acceleration of the universe and the physics behind it}},
   {\em Gen.Rel.Grav.} {\bf 39} (2007) 307--342,
  [\href{http://arxiv.org/abs/astro-ph/0605313}{{\tt astro-ph/0605313}}].

\bibitem{caldwell:07}
R.~Caldwell, A.~Cooray, and A.~Melchiorri, {\it {Constraints on a New
  Post-General Relativity Cosmological Parameter}},  {\em Phys.Rev.} {\bf D76}
  (2007) 023507, [\href{http://arxiv.org/abs/astro-ph/0703375}{{\tt
  astro-ph/0703375}}].

\bibitem{zhang:07}
P.~Zhang, M.~Liguori, R.~Bean, and S.~Dodelson, {\it {Probing Gravity at
  Cosmological Scales by Measurements which Test the Relationship between
  Gravitational Lensing and Matter Overdensity}},  {\em Phys. Rev. Lett.} {\bf
  99} (2007) 141302, [\href{http://arxiv.org/abs/0704.1932}{{\tt
  arXiv:0704.1932}}].

\bibitem{amendola:07}
L.~Amendola, M.~Kunz, and D.~Sapone, {\it {Measuring the dark side (with weak
  lensing)}},  {\em JCAP} {\bf 0804} (2008) 013,
  [\href{http://arxiv.org/abs/0704.2421}{{\tt arXiv:0704.2421}}].

\bibitem{hu:07b}
W.~Hu and I.~Sawicki, {\it {A Parameterized Post-Friedmann Framework for
  Modified Gravity}},  {\em Phys. Rev.} {\bf D76} (2007) 104043,
  [\href{http://arxiv.org/abs/0708.1190}{{\tt arXiv:0708.1190}}].

\bibitem{bertschinger:08}
E.~Bertschinger and P.~Zukin, {\it {Distinguishing Modified Gravity from Dark
  Energy}},  {\em Phys.Rev.} {\bf D78} (2008) 024015,
  [\href{http://arxiv.org/abs/0801.2431}{{\tt arXiv:0801.2431}}].

\bibitem{daniel:10}
S.~F. Daniel et~al., {\it {Testing General Relativity with Current Cosmological
  Data}},  {\em Phys. Rev.} {\bf D81} (2010) 123508,
  [\href{http://arxiv.org/abs/1002.1962}{{\tt arXiv:1002.1962}}].

\bibitem{saltas:14}
I.~D. Saltas, I.~Sawicki, L.~Amendola, and M.~Kunz, {\it {Anisotropic Stress as
  a Signature of Nonstandard Propagation of Gravitational Waves}},  {\em Phys.
  Rev. Lett.} {\bf 113} (2014), no.~19 191101,
  [\href{http://arxiv.org/abs/1406.7139}{{\tt arXiv:1406.7139}}].

\bibitem{gleyzes:14a}
J.~Gleyzes, D.~Langlois, F.~Piazza, and F.~Vernizzi, {\it {Healthy theories
  beyond Horndeski}},  {\em Phys. Rev. Lett.} {\bf 114} (2015), no.~21 211101,
  [\href{http://arxiv.org/abs/1404.6495}{{\tt arXiv:1404.6495}}].

\bibitem{lombriser:15b}
L.~Lombriser and A.~Taylor, {\it {Semi-dynamical perturbations of unified dark
  energy}},  {\em JCAP} {\bf 1511} (2015), no.~11 040,
  [\href{http://arxiv.org/abs/1505.05915}{{\tt arXiv:1505.05915}}].

\bibitem{lombriser:13c}
L.~Lombriser, K.~Koyama, and B.~Li, {\it {Halo modelling in chameleon
  theories}},  {\em JCAP} {\bf 1403} (2014) 021,
  [\href{http://arxiv.org/abs/1312.1292}{{\tt arXiv:1312.1292}}].

\bibitem{lombriser:15a}
L.~Lombriser, F.~Simpson, and A.~Mead, {\it {Unscreening modified gravity in
  the matter power spectrum}},  {\em Phys. Rev. Lett.} {\bf 114} (2015), no.~25
  251101, [\href{http://arxiv.org/abs/1501.04961}{{\tt arXiv:1501.04961}}].

\bibitem{bertotti:03}
B.~Bertotti, L.~Iess, and P.~Tortora, {\it {A test of general relativity using
  radio links with the Cassini spacecraft}},  {\em Nature} {\bf 425} (2003)
  374.

\bibitem{winther:15}
H.~A. Winther et~al., {\it {Modified Gravity N-body Code Comparison Project}},
  {\em Mon. Not. Roy. Astron. Soc.} {\bf 454} (2015), no.~4 4208--4234,
  [\href{http://arxiv.org/abs/1506.06384}{{\tt arXiv:1506.06384}}].

\bibitem{mead:14}
A.~Mead, J.~Peacock, L.~Lombriser, and B.~Li, {\it {Rapid simulation rescaling
  from standard to modified gravity models}},  {\em Mon. Not. Roy. Astron.
  Soc.} {\bf 452} (2015) 4203, [\href{http://arxiv.org/abs/1412.5195}{{\tt
  arXiv:1412.5195}}].

\bibitem{koyama:09}
K.~Koyama, A.~Taruya, and T.~Hiramatsu, {\it {Non-linear Evolution of Matter
  Power Spectrum in Modified Theory of Gravity}},  {\em Phys. Rev.} {\bf D79}
  (2009) 123512, [\href{http://arxiv.org/abs/0902.0618}{{\tt
  arXiv:0902.0618}}].

\bibitem{brax:13}
P.~Brax and P.~Valageas, {\it {Impact on the power spectrum of Screening in
  Modified Gravity Scenarios}},  {\em Phys. Rev.} {\bf D88} (2013), no.~2
  023527, [\href{http://arxiv.org/abs/1305.5647}{{\tt arXiv:1305.5647}}].

\bibitem{schmidt:08}
F.~Schmidt, M.~V. Lima, H.~Oyaizu, and W.~Hu, {\it {Non-linear Evolution of
  f(R) Cosmologies III: Halo Statistics}},  {\em Phys. Rev.} {\bf D79} (2009)
  083518, [\href{http://arxiv.org/abs/0812.0545}{{\tt arXiv:0812.0545}}].

\bibitem{schmidt:09b}
F.~Schmidt, W.~Hu, and M.~Lima, {\it {Spherical Collapse and the Halo Model in
  Braneworld Gravity}},  {\em Phys. Rev.} {\bf D81} (2010) 063005,
  [\href{http://arxiv.org/abs/0911.5178}{{\tt arXiv:0911.5178}}].

\bibitem{lombriser:13b}
L.~Lombriser, B.~Li, K.~Koyama, and G.-B. Zhao, {\it {Modeling halo mass
  functions in chameleon f(R) gravity}},  {\em Phys. Rev.} {\bf D87} (2013),
  no.~12 123511, [\href{http://arxiv.org/abs/1304.6395}{{\tt
  arXiv:1304.6395}}].

\bibitem{lombriser:14a}
L.~Lombriser, {\it {Constraining chameleon models with cosmology}},  {\em
  Annalen Phys.} {\bf 526} (2014) 259--282,
  [\href{http://arxiv.org/abs/1403.4268}{{\tt arXiv:1403.4268}}].

\bibitem{song:06}
Y.-S. Song, W.~Hu, and I.~Sawicki, {\it {The Large Scale Structure of f(R)
  Gravity}},  {\em Phys. Rev.} {\bf D75} (2007) 044004,
  [\href{http://arxiv.org/abs/astro-ph/0610532}{{\tt astro-ph/0610532}}].

\bibitem{giannantonio:09}
T.~Giannantonio, M.~Martinelli, A.~Silvestri, and A.~Melchiorri, {\it {New
  constraints on parametrised modified gravity from correlations of the CMB
  with large scale structure}},  {\em JCAP} {\bf 1004} (2010) 030,
  [\href{http://arxiv.org/abs/0909.2045}{{\tt arXiv:0909.2045}}].

\bibitem{lombriser:10}
L.~Lombriser, A.~Slosar, U.~Seljak, and W.~Hu, {\it {Constraints on f(R)
  gravity from probing the large-scale structure}},  {\em Phys. Rev.} {\bf D85}
  (2012) 124038, [\href{http://arxiv.org/abs/1003.3009}{{\tt
  arXiv:1003.3009}}].

\bibitem{ho:08}
S.~Ho, C.~Hirata, N.~Padmanabhan, U.~Seljak, and N.~Bahcall, {\it {Correlation
  of CMB with large-scale structure: I. ISW Tomography and Cosmological
  Implications}},  {\em Phys. Rev.} {\bf D78} (2008) 043519,
  [\href{http://arxiv.org/abs/0801.0642}{{\tt arXiv:0801.0642}}].

\bibitem{giannantonio:08}
T.~Giannantonio, R.~Scranton, R.~G. Crittenden, R.~C. Nichol, S.~P. Boughn,
  A.~D. Myers, and G.~T. Richards, {\it {Combined analysis of the integrated
  Sachs-Wolfe effect and cosmological implications}},  {\em Phys. Rev.} {\bf
  D77} (2008) 123520, [\href{http://arxiv.org/abs/0801.4380}{{\tt
  arXiv:0801.4380}}].

\bibitem{barreira:14b}
A.~Barreira, B.~Li, C.~Baugh, and S.~Pascoli, {\it {The observational status of
  Galileon gravity after Planck}},  {\em JCAP} {\bf 1408} (2014) 059,
  [\href{http://arxiv.org/abs/1406.0485}{{\tt arXiv:1406.0485}}].

\bibitem{mcmanus:15}
R.~McManus, L.~Lombriser, and J.~Pe\~narrubia. in prep.

\bibitem{yoo:12}
J.~Yoo, N.~Hamaus, U.~Seljak, and M.~Zaldarriaga, {\it {Going beyond the Kaiser
  redshift-space distortion formula: a full general relativistic account of the
  effects and their detectability in galaxy clustering}},  {\em Phys. Rev.}
  {\bf D86} (2012) 063514, [\href{http://arxiv.org/abs/1206.5809}{{\tt
  arXiv:1206.5809}}].

\bibitem{lombriser:13a}
L.~Lombriser, J.~Yoo, and K.~Koyama, {\it {Relativistic effects in galaxy
  clustering in a parametrized post-Friedmann universe}},  {\em Phys.Rev.} {\bf
  D87} (2013) 104019, [\href{http://arxiv.org/abs/1301.3132}{{\tt
  arXiv:1301.3132}}].

\bibitem{seljak:08}
U.~Seljak, {\it {Extracting primordial non-gaussianity without cosmic
  variance}},  {\em Phys.Rev.Lett.} {\bf 102} (2009) 021302,
  [\href{http://arxiv.org/abs/0807.1770}{{\tt arXiv:0807.1770}}].

\bibitem{mcdonald:08}
P.~McDonald and U.~Seljak, {\it {How to measure redshift-space distortions
  without sample variance}},  {\em JCAP} {\bf 0910} (2009) 007,
  [\href{http://arxiv.org/abs/0810.0323}{{\tt arXiv:0810.0323}}].

\bibitem{yoo:09}
J.~Yoo, A.~L. Fitzpatrick, and M.~Zaldarriaga, {\it {A New Perspective on
  Galaxy Clustering as a Cosmological Probe: General Relativistic Effects}},
  {\em Phys. Rev.} {\bf D80} (2009) 083514,
  [\href{http://arxiv.org/abs/0907.0707}{{\tt arXiv:0907.0707}}].

\bibitem{yoo:10}
J.~Yoo, {\it {General Relativistic Description of the Observed Galaxy Power
  Spectrum: Do We Understand What We Measure?}},  {\em Phys. Rev.} {\bf D82}
  (2010) 083508, [\href{http://arxiv.org/abs/1009.3021}{{\tt
  arXiv:1009.3021}}].

\bibitem{bonvin:11}
C.~Bonvin and R.~Durrer, {\it {What galaxy surveys really measure}},  {\em
  Phys. Rev.} {\bf D84} (2011) 063505,
  [\href{http://arxiv.org/abs/1105.5280}{{\tt arXiv:1105.5280}}].

\bibitem{challinor:11}
A.~Challinor and A.~Lewis, {\it {The linear power spectrum of observed source
  number counts}},  {\em Phys. Rev.} {\bf D84} (2011) 043516,
  [\href{http://arxiv.org/abs/1105.5292}{{\tt arXiv:1105.5292}}].

\bibitem{planck13:15}
{\bf Planck} Collaboration, P.~Ade et~al., {\it {Planck 2015 results. XIII.
  Cosmological parameters}},  \href{http://arxiv.org/abs/1502.01589}{{\tt
  arXiv:1502.01589}}.

\bibitem{song:10}
Y.-S. Song, G.-B. Zhao, D.~Bacon, K.~Koyama, R.~C. Nichol, and L.~Pogosian,
  {\it {Complementarity of Weak Lensing and Peculiar Velocity Measurements in
  Testing General Relativity}},  {\em Phys. Rev.} {\bf D84} (2011) 083523,
  [\href{http://arxiv.org/abs/1011.2106}{{\tt arXiv:1011.2106}}].

\bibitem{lombriser:14c}
L.~Lombriser and J.~Pe\~narrubia, {\it {How chameleons core dwarfs with
  cusps}},  {\em Phys.Rev.} {\bf D91} (2015), no.~8 084022,
  [\href{http://arxiv.org/abs/1407.7862}{{\tt arXiv:1407.7862}}].

\bibitem{amendola:14}
L.~Amendola, G.~Ballesteros, and V.~Pettorino, {\it {Effects of modified
  gravity on B-mode polarization}},  {\em Phys. Rev.} {\bf D90} (2014) 043009,
  [\href{http://arxiv.org/abs/1405.7004}{{\tt arXiv:1405.7004}}].

\bibitem{raveri:14}
M.~Raveri, C.~Baccigalupi, A.~Silvestri, and S.-Y. Zhou, {\it {Measuring the
  speed of cosmological gravitational waves}},  {\em Phys. Rev.} {\bf D91}
  (2015), no.~6 061501, [\href{http://arxiv.org/abs/1405.7974}{{\tt
  arXiv:1405.7974}}].

\bibitem{pettorino:14}
V.~Pettorino and L.~Amendola, {\it {Friction in Gravitational Waves: a test for
  early-time modified gravity}},  {\em Phys. Lett.} {\bf B742} (2015) 353--357,
  [\href{http://arxiv.org/abs/1408.2224}{{\tt arXiv:1408.2224}}].

\bibitem{moore:01}
G.~D. Moore and A.~E. Nelson, {\it {Lower bound on the propagation speed of
  gravity from gravitational Cherenkov radiation}},  {\em JHEP} {\bf 09} (2001)
  023, [\href{http://arxiv.org/abs/hep-ph/0106220}{{\tt hep-ph/0106220}}].

\bibitem{caves:80}
C.~M. Caves, {\it {Gravitational radiation and the ultimate speed in Rosen's
  bimetric theory of gravity}},  {\em Annals Phys.} {\bf 125} (1980) 35--52.

\bibitem{aVIRGO}
T.~{Accadia} et~al., {\it {Status of the Virgo project}},  {\em Classical and
  Quantum Gravity} {\bf 28} (June, 2011) 114002.

\bibitem{KAGRA}
K.~{Somiya}, {\it {Detector configuration of KAGRA-the Japanese cryogenic
  gravitational-wave detector}},  {\em Classical and Quantum Gravity} {\bf 29}
  (June, 2012) 124007, [\href{http://arxiv.org/abs/1111.7185}{{\tt
  arXiv:1111.7185}}].

\bibitem{DECIGO}
S.~{Kawamura} et~al., {\it {The Japanese space gravitational wave antenna:
  DECIGO}},  {\em Classical and Quantum Gravity} {\bf 28} (May, 2011) 094011.

\bibitem{eLISA}
P.~{Amaro-Seoane} et~al., {\it {eLISA: Astrophysics and cosmology in the
  millihertz regime}},  {\em GW Notes, Vol.~6, p.~4-110} {\bf 6} (May, 2013)
  4--110, [\href{http://arxiv.org/abs/1201.3621}{{\tt arXiv:1201.3621}}].

\bibitem{BBO}
S.~Phinney et~al., {\it {The Big Bang Observer: Direct detection of
  gravitational waves from the birth of the Universe to the Present}},  {\em
  NASA Mission Concept Study} (2004).

\bibitem{LIGOIII}
R.~X. {Adhikari}, {\it {Gravitational radiation detection with laser
  interferometry}},  {\em Reviews of Modern Physics} {\bf 86} (Jan., 2014)
  121--151, [\href{http://arxiv.org/abs/1305.5188}{{\tt arXiv:1305.5188}}].

\bibitem{ET}
M.~{Punturo} et~al., {\it {The Einstein Telescope: a third-generation
  gravitational wave observatory}},  {\em Classical and Quantum Gravity} {\bf
  27} (Oct., 2010) 194002.

\bibitem{nishizawa:14}
A.~Nishizawa and T.~Nakamura, {\it {Measuring Speed of Gravitational Waves by
  Observations of Photons and Neutrinos from Compact Binary Mergers and
  Supernovae}},  {\em Phys. Rev.} {\bf D90} (2014), no.~4 044048,
  [\href{http://arxiv.org/abs/1406.5544}{{\tt arXiv:1406.5544}}].

\bibitem{cutler:09}
C.~Cutler and D.~E. Holz, {\it {Ultra-high precision cosmology from
  gravitational waves}},  {\em Phys. Rev.} {\bf D80} (2009) 104009,
  [\href{http://arxiv.org/abs/0906.3752}{{\tt arXiv:0906.3752}}].

\bibitem{abe:11}
K.~Abe et~al., {\it {Letter of Intent: The Hyper-Kamiokande Experiment ---
  Detector Design and Physics Potential ---}},
  \href{http://arxiv.org/abs/1109.3262}{{\tt arXiv:1109.3262}}.

\bibitem{schutz:86}
B.~F. Schutz, {\it {Determining the Hubble Constant from Gravitational Wave
  Observations}},  {\em Nature} {\bf 323} (1986) 310--311.

\bibitem{holz:05}
D.~E. Holz and S.~A. Hughes, {\it {Using gravitational-wave standard sirens}},
  {\em Astrophys. J.} {\bf 629} (2005) 15--22,
  [\href{http://arxiv.org/abs/astro-ph/0504616}{{\tt astro-ph/0504616}}].

\bibitem{shapiro:09}
C.~Shapiro, D.~Bacon, M.~Hendry, and B.~Hoyle, {\it {Delensing Gravitational
  Wave Standard Sirens with Shear and Flexion Maps}},  {\em Mon. Not. Roy.
  Astron. Soc.} {\bf 404} (2010) 858--866,
  [\href{http://arxiv.org/abs/0907.3635}{{\tt arXiv:0907.3635}}].

\end{thebibliography}\endgroup

\end{document}